\documentclass[twocolumn,twocolappendix,tighten,times]{aastex631}

\usepackage{booktabs}
\usepackage{graphicx}
\usepackage{amsmath}
\usepackage{verbatim}

\submitjournal{ApJ}

\begin{document}

\title{Testing the Bullet Dwarf Collision Scenario in the NGC 1052 Group Through Morphologies and Stellar Populations}
\shorttitle{NGC 1052 Trail Dwarfs}
\shortauthors{Tang et al.}


\correspondingauthor{Yimeng Tang}
\email{ymtang@ucsc.edu}

\author{Yimeng Tang}
\affiliation{Department of Astronomy \& Astrophysics, University of California Santa Cruz, 1156 High Street, Santa Cruz, CA 95064, USA}

\author{Aaron J.\ Romanowsky}
\affiliation{Department of Astronomy \& Astrophysics, University of California Santa Cruz, 1156 High Street, Santa Cruz, CA 95064, USA}
\affiliation{Department of Physics \& Astronomy, San Jos\'e State University, One Washington Square, San Jose, CA 95192, USA}

\author{Pieter G.\ van Dokkum}
\affiliation{Department of Astronomy, Yale University, New Haven, CT 06511, USA}

\author{T.\ H.\ Jarrett}
\altaffiliation{Deceased.}
\affiliation{Department of Astronomy, University of Cape Town, Rondebosch, 7700, South Africa}

\author{Kevin A.\ Bundy}
\affiliation{Department of Astronomy \& Astrophysics, University of California Santa Cruz, 1156 High Street, Santa Cruz, CA 95064, USA}

\author{Maria Luisa Buzzo}
\affiliation{Centre for Astrophysics and Supercomputing, Swinburne University, John Street, Hawthorn, VIC 3122, Australia}
\affiliation{ARC Centre of Excellence for All Sky Astrophysics in 3 Dimensions (ASTRO 3D), Australia}

\author{Shany Danieli}
\affiliation{Department of Astrophysical Sciences, Princeton University, Princeton, NJ 08544, USA}

\author{Jonah S. Gannon}
\affiliation{Centre for Astrophysics and Supercomputing, Swinburne University, John Street, Hawthorn, VIC 3122, Australia}
\affiliation{ARC Centre of Excellence for All Sky Astrophysics in 3 Dimensions (ASTRO 3D), Australia}

\author{Michael A.\ Keim}
\affiliation{Department of Astronomy, Yale University, New Haven, CT 06511, USA}

\author{Seppo Laine}
\affiliation{IPAC, Mail Code 314-6, Caltech, 1200 E. California Blvd., Pasadena, CA 91125, USA}

\author{Zili Shen}
\affiliation{Department of Astronomy, Yale University, New Haven, CT 06511, USA}

\begin{abstract}

NGC~1052-DF2 and -DF4 are two ultra-diffuse galaxies that have been reported as deficient in dark matter and associated with the same galaxy group. Recent findings suggest that DF2 and DF4 are part of a large linear substructure of dwarf galaxies that could have been formed from a high-velocity head-on encounter of two gas-rich galaxies, known as a bullet dwarf collision. Based on new observations from the {\it Hubble Space Telescope}, combined with existing imaging from the $u$ band to mid-infrared, we test the bullet dwarf scenario by studying the morphologies and stellar populations of the trail dwarfs. We find no significant morphological differences between the trail dwarfs and other dwarfs in the group, while for both populations, their photometric major axes unexpectedly align parallel with the trail. We find that the trail dwarfs have significantly older ages and higher metallicities than the comparison sample, supporting the distinctiveness of the trail. These observations provide key constraints for any formation model, and we argue that they are currently best explained by the bullet dwarf collision scenario, with additional strong tests anticipated with future observations.

\end{abstract}


\section{Introduction}
\label{intro}

Galaxies are understood to form under the influence of their dark matter (DM) haloes, with dwarf galaxies among the most DM-dominated systems, owing to their susceptibility to internal feedback. It was therefore surprising when two nearby dwarfs, NGC~1052-DF2 and NGC~1052-DF4 (hereafter DF2 and DF4), were reported to contain little if any DM \citep{vanDokkum2018a,vanDokkum2019}. The primary evidence is from the low velocity dispersions of the galaxies' stars and globular cluster (GC) systems \citep{Wasserman2018,Danieli2019,Emsellem2019,Shen2023}, and is supported by strong distortions in their outer isophotes, suggesting sensitivity to tidal effects in the absence of DM \citep{Keim2022}. Various questions have been raised about the mass inferences (e.g., \citealt{Trujillo2019}), but the key concern about the line-of-sight distance has been settled through observations of the tip of the red giant branch (TRGB) using the {\it Hubble Space Telescope} ({\it HST}; \citealt{vanDokkum2018c,Danieli2020,Shen2021b}). 

Attention has now shifted to understanding the formation mechanisms of these intriguing galaxies. Scenarios where galaxies are formed without DM haloes include tidal dwarfs formed from gas thrown out from interactions between massive galaxies or from ram-pressure stripping \citep{Fensch2019a,Lora2024}, and supermassive black hole jet- or outflow-induced star formation (e.g., \citealt{vanBreugel1985,Natarajan1998,Zovaro2020}). In other scenarios, the dwarfs formed normally but then lost their DM, either through tidal stripping (e.g., \citealt{Ogiya2018,Carleton2019,Nusser2020,Montes2020,Ogiya2022,Moreno2022,Katayama2023,Golini2024,Montero-Dorta2024}) or through extreme feedback that propels the DM out of the central regions \citep{TrujilloGomez2021,TrujilloGomez2022}.

Important clues for discriminating among formation models are provided by other properties of these dwarfs besides their masses. Their star clusters are larger and much more luminous than typical GCs \citep{vanDokkum2018b,Ma2020,Shen2021a}. The GCs are also monochromatic within and between DF2 and DF4, implying unusually synchronized single-burst formation histories \citep{vanDokkum2022b}. The two galaxies themselves also have similar ages, metallicities \citep{Buzzo2022} and morphologies \citep{Keim2022}, and overall there is a strong impression of a shared formation history. On the other hand, the TRGB distance remarkably shows that DF2 and DF4 are widely separated by $\sim$~2~Mpc \citep{Shen2021b,Shen2023}, and therefore one or both of them are {\it not} currently in the NGC~1052 group as initially assumed from their projected sky positions.

These additional observations appear difficult to explain in any of the formation scenarios mentioned above. So far the only proposed scenario that shows potential for fitting all the constraints is a ``bullet dwarf'' event where two gas-rich galaxies collided at high velocity in a proto-group environment \citep{Silk2019,Shin2020,Lee2021,Otaki2023,Lee2024}. The gas would be shocked and separated from the DM in the collision, creating one or more diffuse, DM-free galaxies between the two progenitor DM haloes. Unusually luminous GCs could form through high compression of clumps within the gas. This scenario nicely explains the existence of {\it two} galaxies in the same system, DF2 and DF4, with similarly unusual properties, while the high velocities make it plausible that one or both are unbound to the central galaxy NGC~1052. Furthermore, the spectroscopic age estimates of stars and GCs (7--11 Gyr; \citealt{vanDokkum2018b,Fensch2019a}) agree with a backward extrapolation of the galaxies' trajectories to a common origin point based on relative line-of-sight distance and velocity ($\sim$~6--8 Gyr; \citealt{vanDokkum2022a}).

Simulations also suggest that a bullet dwarf collision can create three or more DM-free dwarfs, leading to an apparent trail of galaxies \citep{Shin2020}. \citet{vanDokkum2022a} analyzed the spatial distribution of dwarfs around NGC~1052, and found a significant linear overdensity of around ten galaxies that includes DF2 and DF4, with dimensions of $\sim 70\times10$~arcmin ($\sim 400 \times 60$~kpc). This feature provided new and unique evidence for the bullet dwarf scenario, since there is no natural expectation for any of the other DF2/DF4 formation scenarios to lead to a trail of dwarfs. Additional simulations were able to reproduce the observed trail dwarf positions from a bullet dwarf event, under specific initial conditions \citep{Lee2024}.

It is now crucial to examine the ``trail dwarfs'' in more detail, to confirm that they truly comprise a physical association, and to test predictions from the bullet dwarf scenario. If these dwarfs were formed together in a single event, then they should have relatively homogeneous properties that are distinct from those of other dwarfs around NGC~1052. We will defer study of some of these properties to future work: DM content, line-of-sight distances, and redshifts. Here we focus on the morphologies and stellar populations of the dwarfs, making use of spectroscopic data as well as multi-wavelength imaging that includes a fresh set of observations from {\it HST}. We note that \cite{Buzzo2023} investigated the GC systems of the dwarfs from ground-based imaging, and we expect that the {\it HST} imaging will provide stronger constraints in future work.

The rest of this paper is structured as follows. In Sections \ref{data} and \ref{methods}, we describe our data and data analysis methods. The main results and discussion are given in Sections \ref{results_morph} (morphologies), \ref{results_cmr} (color--magnitude relations) and \ref{results_sed} (stellar populations). We compare different formation theories in Section \ref{discussion_theories}. A summary follows in Section \ref{conclusions}. More details and tests of the morphology and stellar population methods and results are presented in Appendices \ref{morph_appendix} and \ref{stellar_pop_appendix}, respectively. A default distance of $20.4 \pm 1.0$~Mpc is adopted based on NGC~1052 \citep{Blakeslee2001,Tonry2001,Tully2013}, with variations about this distance discussed when relevant (note the mean distance to DF2/DF4 is $21.0\pm1.0$~Mpc; \citealt{Shen2023}).

\section{Data}
\label{data}

\begin{figure*}
\centering
\includegraphics[width=1\textwidth]{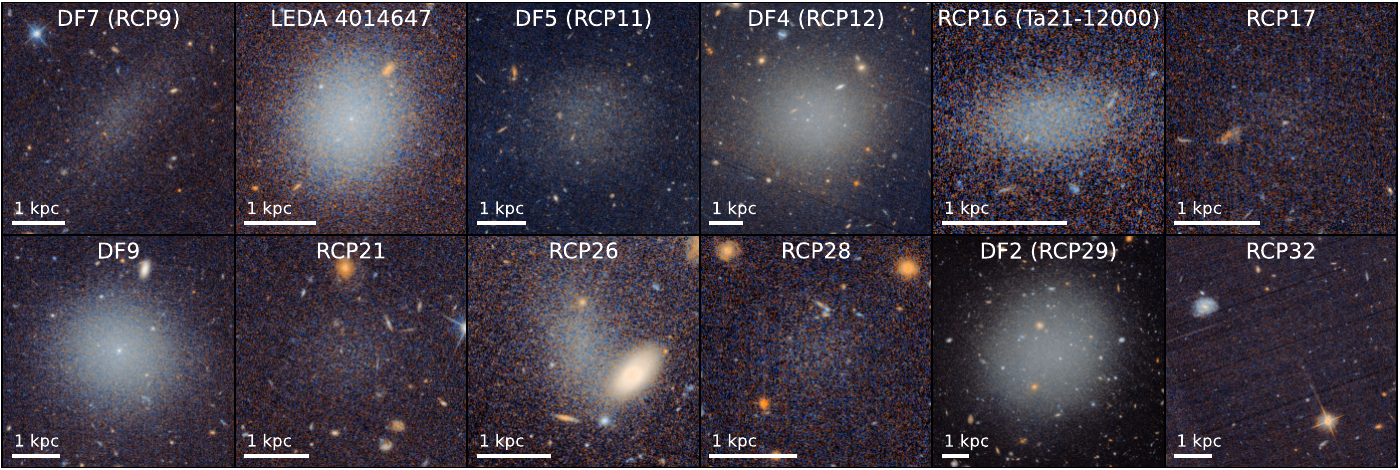}
\caption{The image gallery of dwarf galaxies on the trail structure. The galaxies are ordered by their right ascension from left to right and top to bottom. The pseudo-color images are created using the $V_{606}$ and $I_{814}$ bands from {\it HST}, with the \citet{Lupton2004} algorithm. The cutout image size is approximately 5 times the circularized effective radius of each galaxy, and 1 kpc scale-bars indicate physical sizes. North is up and East is left.}
\label{fig:trail_gallery}
\end{figure*}

\begin{figure*}
\centering
\includegraphics[width=1\textwidth]{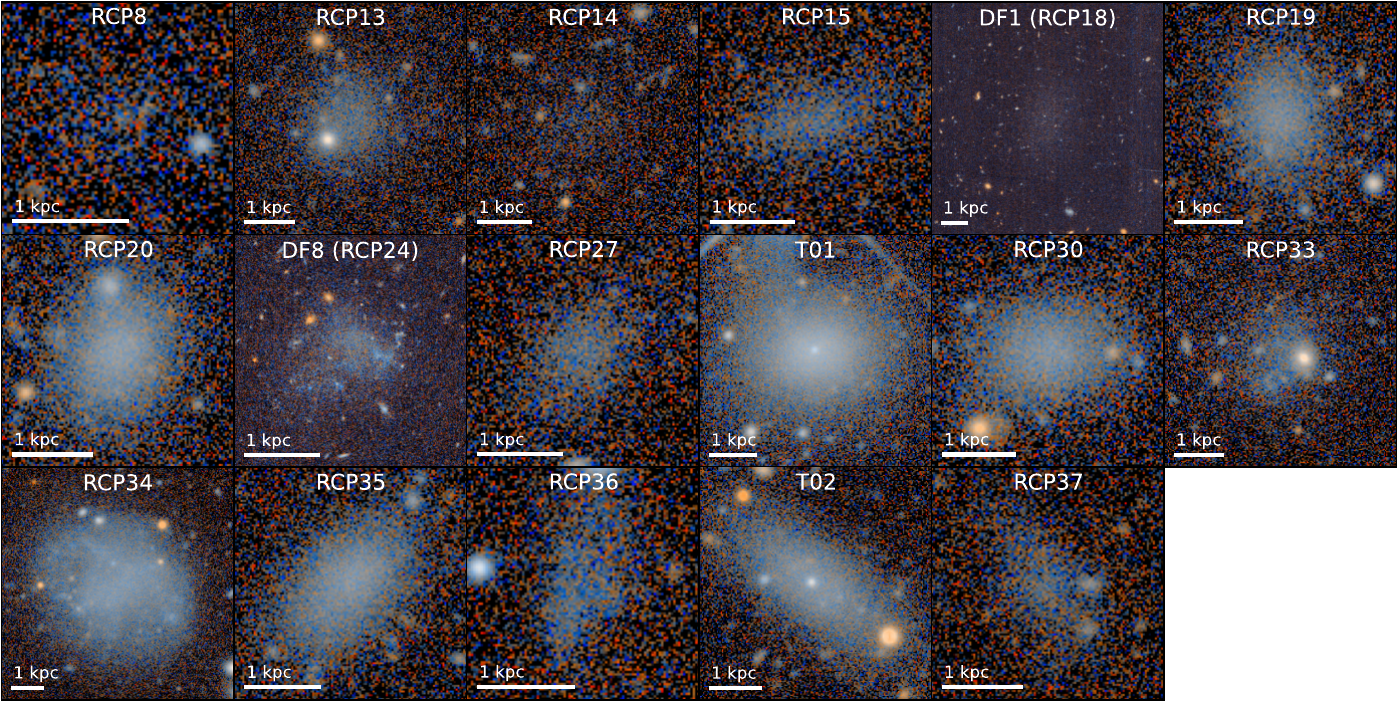}
\caption{The image gallery of dwarf galaxies near NGC~1052 in projection and {\it not} associated with the trail structure. See Figure \ref{fig:trail_gallery} for details, except here only two galaxies are imaged with {\it HST} (DF1 and DF8), while the rest use the $r$ and $i$ bands from DECaLS.}
\label{fig:nontrail_gallery}
\end{figure*}

In this section, we discuss the galaxy sample selection (Section~\ref{sec:sample} and provide details about the imaging data used in this work (Section \ref{sec:imaging}). Some complementary spectroscopy is discussed in Appendix~\ref{stellar_pop_appendix}.

\subsection{Galaxy sample selection}\label{sec:sample}

In this paper we focus on a circular region with 1-degree radius from the central group galaxy NGC~1052, extending slightly beyond the region spanned by the trail. We find that the number density of dwarf galaxies drops sharply beyond this radius, which corresponds to 360~kpc at the distance of the group, and is close to the estimated virial radius \citep{Forbes2019}. We select dwarf galaxies ($M_* \lesssim 10^9 M_{\odot}$) with total apparent $g$-band magnitudes\footnote{All magnitudes in this work are in the AB system.} in the range $g_0>15.5$. These include the 12 candidate trail dwarfs identified by \cite{vanDokkum2022a}. We note that our trail dwarfs sample includes DF9, which was not used in the main analysis of low surface brightness (LSB) dwarfs in \cite{vanDokkum2022a}.

Apart from trail dwarfs, all other dwarfs selected are considered a ``non-trail" comparison sample. The non-trail dwarfs are initially drawn from the study of LSB galaxies by \cite{Roman2021}, which also included almost all of the trail dwarfs. We have also checked the LSB galaxies identified by \cite{Tanoglidis2021} and \cite{Trujillo2021} in the same sky region to search for any potentially missed group members fitting our selection criteria. We found one candidate near the trail, Ta21-11818 from \citet{Tanoglidis2021} (024132$-$081745 from \citealt{Paudel2023}). This galaxy was excluded by \citet{Roman2021} because its small size could mean it is a background object, and we therefore also exclude it from our sample. \citet{Paudel2023} reported two additional candidates that are relatively bright, SDSS J024117.24$-$075356.8 and SDSS J024321.87$-$075032.7, which they called 024117$-$075356 and 024321$-$075032 and which we designate for simplicity as NGC~1052-T01 and -T02 (T02 is also SMDG0243218$-$075033 in \citealt{Zaritsky2022}). This gives us a sample of 17 non-trail dwarf candidates (keeping in mind that redshifts or other distance confirmations are not available for most of them). Pseudo-color images of trail dwarfs and non-trail dwarfs are shown in Figures \ref{fig:trail_gallery} and \ref{fig:nontrail_gallery}. The complete list of dwarfs in our sample can be found in Table \ref{tab:data_properties}.

\begin{deluxetable*}{lcccccccc}
\tabletypesize{\footnotesize}
\renewcommand{\arraystretch}{1.05}
\tablewidth{0pt}
\tablecaption{Sky positions and morphological properties of the dwarf galaxies in our sample, sorted by right ascension. {\it HST} $V_{606}+I_{814}$ stacked images are preferred for measurement, and if not available, we use DECaLS $g+r$ stacked images instead. All properties are obtained from GALFIT, including major-axis effective radius $R_{\rm e}$, S\'ersic index $n$, axis ratio {\it b/a}, position angle and extinction-corrected mean $g$-band surface brightness within the effective radius. The fourth column indicates whether or not the galaxy is part of the trail (based on sky positions).
\label{tab:data_properties}}
\tablehead{
\colhead{Galaxy} & \colhead{RA} & \colhead{Dec} & \colhead{On trail?} & \colhead{$R_{\rm e}$} & \colhead{$n$} & \colhead{{\it b/a}} & \colhead{P.A.} & \colhead{$\langle \mu_{g} \rangle_{\rm e}$} \\
\colhead{} & \colhead{[deg]} & \colhead{[deg]} & \colhead{} & \colhead{[arcsec]} & \colhead{} & \colhead{} & \colhead{[deg]} & \colhead{${\rm mag/arcsec^2}$}}
\startdata
    RCP 8 & 39.59835 & $-$8.22162 & No & $5.75 \pm 2.32$ & $1.221 \pm 0.373$ & $0.537 \pm 0.188$ & $-87.8 \pm 36.3$ & $26.91 \pm 0.70 $\\
    DF7 (RCP 9) & 39.62401 & $-$7.92576 & Yes & $14.66 \pm 1.14$ & $0.798 \pm 0.064$ & $0.431 \pm 0.015$ & $-39.4 \pm 1.1$ & $26.47 \pm 0.17$\\
    LEDA 4014647 & 39.70206 & $-$8.04938 & Yes & $7.53 \pm 0.05$ & $0.703 \pm 0.009$ & $0.866 \pm 0.006$ & $-20.5 \pm 1.4$ &       $23.80 \pm 0.02$\\
    DF5 (RCP 11) & 39.80284 & $-$8.14083 & Yes & $11.85 \pm 0.20$ & $0.549 \pm 0.018$ & $0.782 \pm 0.011$ & $-79.4 \pm 1.9$ & $26.32 \pm 0.13$\\
    DF4 (RCP 12) & 39.81271 & $-$8.11597 & Yes & $17.10 \pm 0.07$ & $0.825 \pm 0.005$ & $0.863 \pm 0.003$ & $-77.9 \pm 0.6$ & $24.95 \pm 0.02$\\
    RCP 13 & 39.82676 & $-$7.53675 & No & $10.44 \pm 1.11$ & $0.762 \pm 0.147$ & $0.938 \pm 0.058$ & $-41.4 \pm 36.4$ & $26.08 \pm 0.13$\\
    RCP 14 & 39.86162 & $-$7.37096 & No & $10.53 \pm 2.58$ & $0.407 \pm 0.243$ & $0.768 \pm 0.178$ & $-32.6 \pm 31.0$ & $27.33 \pm 0.47 $\\
    RCP 15 & 39.91021 & $-$7.47356 & No & $8.24 \pm 1.01$ & $0.658 \pm 0.165$ & $0.501 \pm 0.074$ & $-77.1 \pm 5.3$ &         $26.06 \pm 0.18$\\
    RCP 16 (Ta21-12000) & 39.91385 & $-$8.22845 & Yes & $5.25 \pm 0.16$ & $0.665 \pm 0.031$ & $0.543 \pm 0.012$ & $-84.7 \pm 1.0$ & $24.78 \pm 0.09$\\
    RCP 17 & 39.96969 & $-$8.21206 & Yes & $6.22 \pm 0.80$ & $0.640 \pm 0.199$ & $0.859 \pm 0.102$ & $60.2 \pm 33.1$ & $27.14 \pm 0.45$\\
    DF1 (RCP 18) & 40.01907 & $-$8.44619 & No & $26.50 \pm 4.52$ & $1.042 \pm 0.328$ & $0.616 \pm 0.111$ & $-6.7 \pm 15.9$ & $27.44 \pm 0.16$\\
    DF9 & 40.02927 & $-$8.22902 & Yes & $11.38 \pm 0.10$ & $0.819 \pm 0.012$ & $0.884 \pm 0.005$ & $70.1 \pm 1.2$ & $24.49 \pm 0.02$\\
    RCP 19 & 40.03401 & $-$7.94724 & No & $8.45 \pm 0.66$ & $0.843 \pm 0.096$ & $0.743 \pm 0.043$ & $3.7 \pm 6.0$ & $25.28 \pm 0.09$\\
    RCP 20 & 40.08191 & $-$7.98464 & No & $6.87 \pm 0.35$ & $0.823 \pm 0.077$ & $0.801 \pm 0.038$ & $-22.4 \pm 6.7$ & $24.70 \pm 0.07$\\
    RCP 21& 40.12004 & $-$8.24357 & Yes & $8.91 \pm 0.82$ & $0.608 \pm 0.079$ & $0.775 \pm 0.055$ & $-82.3 \pm 9.1$ & $26.60 \pm 0.19$\\
    DF8 (RCP 24) & 40.18932 & $-$7.64687 & No & $7.05 \pm 0.18$ & $0.692 \pm 0.041$ & $0.855 \pm 0.022$ & $-72.9 \pm 5.7$ & $25.31 \pm 0.06$\\
    RCP 26 & 40.28970 & $-$8.29691 & Yes & $8.46 \pm 0.20$ & $0.719 \pm 0.032$ & $0.632 \pm 0.015$ & $17.7 \pm 1.5$ & $25.28 \pm 0.07$\\
    RCP 27 & 40.31254 & $-$7.49335 & No & $7.24 \pm 1.58$ & $0.798 \pm 0.255$ & $0.642 \pm 0.103$ & $-36.9 \pm 10.8$ & $26.04 \pm 0.23$\\
    T01 & 40.32179 & $-$7.89914 & No & $11.96 \pm 0.71$ & $0.968 \pm 0.058$ & $0.810 \pm 0.016$ & $89.2 \pm 2.5$ & $24.27 \pm 0.05$\\
    RCP 28 & 40.42152 & $-$8.34742 & Yes & $6.15 \pm 0.62$ & $0.622 \pm 0.101$ & $0.834 \pm 0.061$ & $-45.4 \pm 12.1$ & $26.60 \pm 0.32$\\
    DF2 (RCP 29)& 40.44531 & $-$8.40297 & Yes & $21.49 \pm 0.03$ & $0.601 \pm 0.002$ & $0.890\pm 0.002$ & $-49.6 \pm 0.4$ & $25.08 \pm 0.02$\\
    RCP 30 & 40.44745 & $-$8.78533 & No & $7.87 \pm 0.45$ & $0.724 \pm 0.078$ & $0.740 \pm 0.040$ & $-82.7 \pm 4.3$ & $24.98 \pm 0.07$\\
    RCP 32 & 40.62034 & $-$8.37512 & Yes & $15.57 \pm 1.86$ & $0.410 \pm 0.256$ & $0.812 \pm 0.166$ & $73.4 \pm 27.4$ & $28.20 \pm 0.38$\\
    RCP 33 & 40.65042 & $-$8.04230 & No & $10.75 \pm 1.10$ & $ 0.720 \pm 0.148$ & $0.900 \pm 0.070$ & $-43.4 \pm 26.0$ & $26.03 \pm 0.15$\\
    RCP 34 & 40.65826 & $-$7.33803 & No & $17.32 \pm 0.20$ & $ 0.519 \pm 0.013$ & $0.884 \pm 0.012$ & $75.4 \pm 3.3$ & $24.37 \pm 0.01$\\
    RCP 35 & 40.69626 & $-$7.77206 & No & $8.39 \pm 0.43$ & $ 0.768 \pm 0.063$ & $0.606 \pm 0.027$ & $-43.8 \pm 2.5$ & $24.90 \pm 0.06$\\
    RCP 36 & 40.76355 & $-$8.01395 & No & $7.04 \pm 1.02$ & $ 0.591 \pm 0.182$ & $0.525 \pm 0.080$ & $-13.5 \pm 5.7$ & $25.92 \pm 0.16$\\
    T02 & 40.84088 & $-$7.84242 & No & $14.79 \pm 0.50$ & $ 0.787 \pm 0.045$ & $0.437 \pm 0.013$ & $53.2 \pm 1.1$ & $24.95 \pm 0.04$\\
    RCP 37 & 40.87002 & $-$7.87319 & No & $8.18 \pm 1.22$ & $ 0.674 \pm 0.228$ & $0.564 \pm 0.088$ & $36.2 \pm 7.5$ & $26.39 \pm 0.18$\\
\enddata
\end{deluxetable*}

\subsection{Photometric data}\label{sec:imaging}

We use photometric data from the optical to mid-infrared (mid-IR) as follows:

\begin{itemize}

\item 
{\it HST} ACS/WFC images in the F606W and F814W bands, hereafter called $V_{606}$ and $I_{814}$ for short. These were obtained in the programs GO 14644, 15695, 15851 and 16912 (PI: Pieter van Dokkum). Program 16912 specifically targeted the previously unobserved trail dwarfs, with observations in Jul and Sep 2022, and these data are presented here for the study of trail dwarfs for the first time. From our sample (see previous section), 14 galaxies were observed by {\it HST}: 12 trail dwarfs and 2 non-trail dwarfs (DF1 and DF8). All of these galaxies have observations of one orbit each in $V_{606}$ and $I_{814}$, except DF2 (19+19 orbits), DF4 (3+7 orbits) and DF5 (3+7 orbits). For all programs, we carried out a special flat-fielding procedure for ACS, applying corrections to the \texttt{flc} files prior to drizzling, as described in the most recent analyses of DF2 and DF4 by \cite{vanDokkum2022b}. We also make small zero-point magnitude adjustments as described in Section \ref{zpcorr}. All the {\it HST} data used in this paper can be found in MAST: \dataset[10.17909/7mt8-8y97]{http://dx.doi.org/10.17909/7mt8-8y97}.

\item Canada--France--Hawaii Telescope (CFHT) MegaCam archival data observed on 2020 September 12 in the $u$, $g$ and $i$ bands (Program ID: 20BO44), presented in \cite{Buzzo2023}. The $1\times1$~degree field of view covers the core regions of the group, including all of the trail dwarfs from our sample except for DF7, and five of the non-trail dwarfs. The total exposure time per band is 11880, 1675 and 2275 seconds, and the average seeing is 0.96, 0.80 and 0.76 arcsec, respectively. Additional information about the data processing can be found in \cite{Buzzo2023}. We also note a systematic magnitude difference of about 0.1 mag in all three bands when comparing stars in the CFHT images and SDSS archive, and therefore applied such corrections to our photometry (see Section \ref{zpcorr}).

\item $g$, $r$, $i$ and $z$ band images from the Dark Energy Camera Legacy Survey (DECaLS; \citealt{Dey2019}). This is the same data source used by \citet{Roman2021} for their dwarf inventory, but is a significantly deeper version (Data Release 10 versus 7), and also has an addition of $i$ band. The DECaLS imaging provides the only optical data with complete coverage of our sample of dwarfs.

\item {\it Spitzer} Infrared Array Camera (IRAC) data in the 3.6$\mu$m band from the program GO 14114 (PI: A.\ Romanowsky), presented also in \citet{Buzzo2022}. Only DF2, DF4 and DF5 were targeted, yielding deeper and higher-resolution imaging than the {\it WISE} W1 data (below), and providing a sanity check for the {\it WISE} photometry that is used for the entire sample.

\item W1, W2, W3, W4 bands from the {\it Wide-field Infrared Survey Explorer} ({\it WISE}; \citealt{Wright2010}), with effective wavelengths of 3.4, 4.6, 12.1 and 22.2 $\mu$m. We use the {\tt ICORE} algorithm \citep{Masci2009,Jarrett2012} to create coadded images based on all {\it WISE} exposures within a 2-degree diameter region centered on NGC~1052. These images were generated at a spatial sampling of 1 arcsec per pixel, which is higher resolution than the publicly available unWISE data (2.75 arcsec per pixel). This super-sampling of the point spread function (PSF), with its full-width at half maximum  (FWHM) of $\sim$~6 arcsec in the W1 and W2 bands, is helpful in modeling and subtracting contaminants. The images are slightly different from those used in \citet{Buzzo2022}, as here there is no removal of stars and other contaminants until the GALFIT modeling stage (below). The {\it WISE} imaging is remarkably deep enough to study LSB dwarf galaxies, and the inclusion of infrared photometry is a key driver of the feasibility of this project.

\end{itemize}

Note that although {\it Galaxy Evolution Explorer} ({\it GALEX}; \citealt{Martin2005}) imaging in the near-ultraviolet is also available, there is little to no detection for most of the galaxies in our sample, and adding these data would do little to improve the stellar population constraints.

The typical SB depths are 29.9 and 29.5 mag arcsec$^{-2}$ in the {\it HST} $V_{606}$ and $I_{814}$ filters with one orbit observation, and the images of DF2, DF4 and DF5 are even deeper. For the DECaLS imaging, the depths are 29.2, 28.8, 27.6 and 28.3 mag  arcsec$^{-2}$ in the $g$, $r$, $i$ and $z$ filters. All the depths are measured as 3$\sigma$ in $10\times10$ arcsec boxes following the depth definition by \cite{Roman2020}. Therefore, we prioritize using the deeper {\it HST} imaging to model the SB distributions of our galaxies.

\section{Methods}
\label{methods}

In this section, we discuss the methods used for analyzing the imaging data, including fitting the SB distributions of the galaxies (Section~\ref{sec:galfit}), zero-point recalibration (Section~\ref{sec:recal}) and spectral energy distribution (SED) fitting (Section~\ref{sec:prosp}).

\subsection{GALFIT Photometry}\label{sec:galfit}
\label{galfit}

\begin{figure*}
\centering
\includegraphics[width=0.6\textwidth]{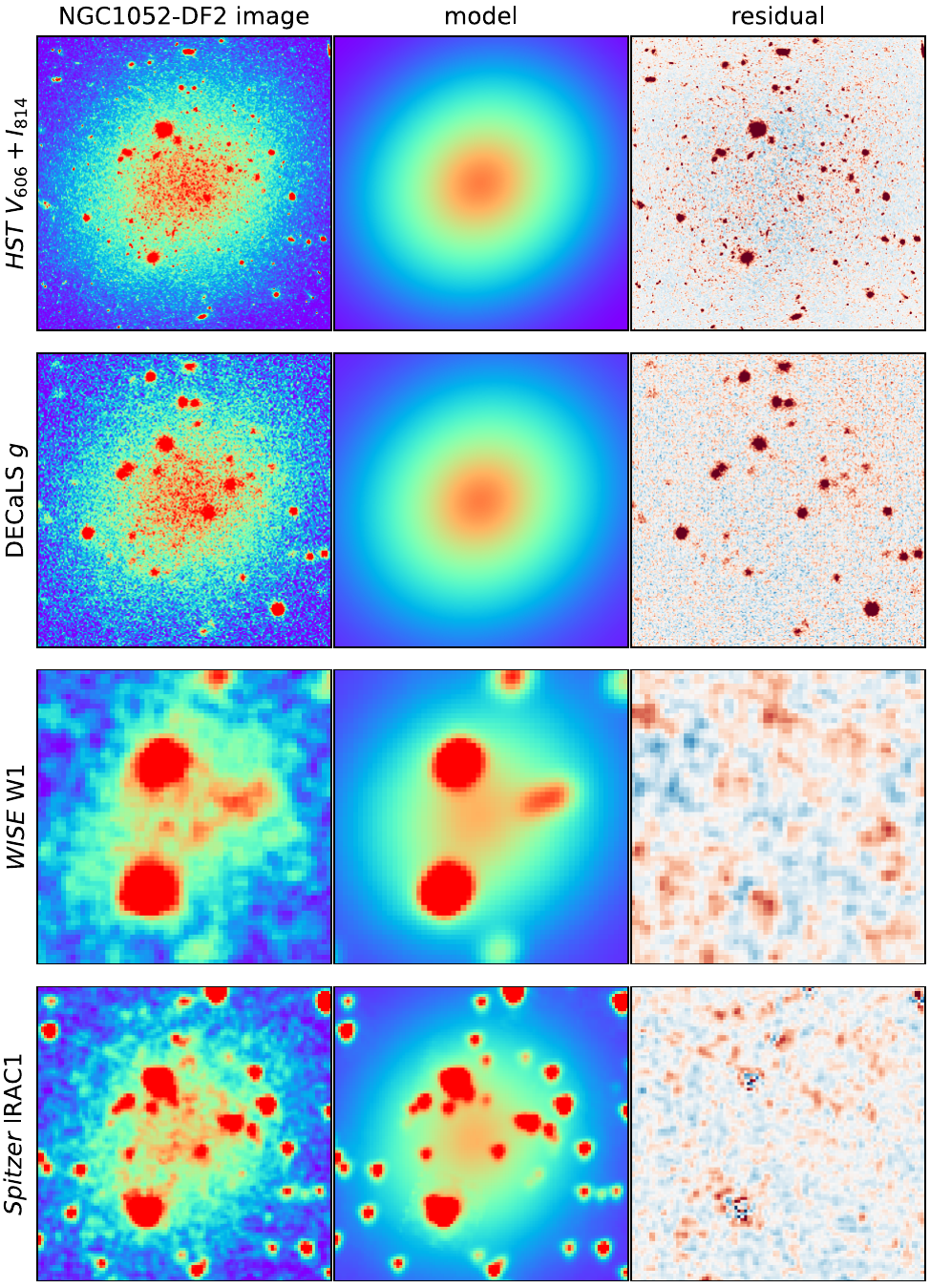}
\caption{DF2 is shown as an example of running GALFIT on multiple wavelength bands from different telescopes. From top to bottom, we present the fitting results for the {\it HST} $V_{606}+I_{814}$ stacked image, DECaLS $g$-band image, {\it WISE} W1 image, and {\it Spitzer} IRAC1 3.6 $\mu$m image of DF2. In each row, from left to right, the three panels show the original image, model and residuals. The images shown here have sizes of 3 times the circularized $R_{\rm e}$ of DF2, while the actual input images to GALFIT typically have a side length of about 8~$R_{\rm e}$. We first fit the {\it HST} $V_{606}+I_{814}$ stacked image to obtain the morphological parameters (if {\it HST} observations are not available, the fitting starts with the DECaLS $g+r$ stacked image). Then these morphological parameters are applied to all other images to get the magnitudes. For infrared imaging, the contamination sources are fitted simultaneously using PSF models to ensure better accuracy.}
\label{fig:galfit_example}
\end{figure*}

We use GALFIT \citep{Peng2002} to perform two-dimensional (2D) galaxy modelling to derive the structural properties, magnitudes and colors of the galaxies. First, for each galaxy, we fit the structural parameters using the image with the highest signal-to-noise ratio, specifically the stacked image of two {\it HST} bands for all the galaxies with {\it HST} observations, and the stacked image of DECaLS with the $g$ and $r$ bands for the other galaxies where {\it HST} imaging is not available. The widths of the cutout images for fitting are typically eight times the effective radius $R_{\rm e}$ of each galaxy. We run SExtractor \citep{Bertin1996} with a low threshold to generate our masks, aiming to minimize contamination from foreground stars and background galaxies. We fit a single S\'ersic model to the SB distribution of each dwarf galaxy, while adding a plane sky model for background estimation. Such a combination of models generally provides reasonable fits to our data. \citet{Buzzo2022} fitted the SB distribution of DF2 and DF4 with a double S\'ersic function, but we find that this makes little difference to the total magnitudes and colors. For the galaxies with nuclei, we mask this component when {\it HST} imaging is used, while for the non-{\it HST} images, we add an extra, central PSF component in the modeling (nearby point sources are used to construct the PSF). All parameters are allowed to vary, but we iteratively run GALFIT and update the initial guess for model parameters, sizes of cutout images and masks in each run until the results are stable. We tested using larger cutout images at the beginning, or changing the threshold for generating the masks, and found no significant differences in the SED shape of galaxies in our sample. For RCP~17 and RCP~32 with extremely low SB, we apply the rebinning method described in \cite{Roman2021} to obtain more reliable measurements from their {\it HST} images. To be more specific, we rebin the images with $4 \times 4$ and $40 \times 40$ pixels (or $0.2 \times 0.2''$ and $2 \times 2''$) for RCP~17 and RCP~32, respectively.

Once the structural parameters are determined ($R_\mathrm{e}$, S\'ersic index $n$, axis ratio $b/a$, position angle; see Table~\ref{tab:data_properties} and example fits in Figure~\ref{fig:galfit_example}), we fix their values and apply them to every single image of each galaxy, leaving only the magnitude and background model free. Here we assume that these galaxies do not have color gradients at a level that would complicate our analysis, which we confirmed by measuring color profiles for the brighter galaxies (DF2, DF4, DF9 and LEDA 4014647) in our sample (see Appendix \ref{morph_appendix}). For DF2, \cite{Fensch2019a} found a metallicity gradient consistent with zero, and \cite{Golini2024} obtained a non-zero but very tiny $g-r$ color gradient which has a negligible effect on the photometric fitting. The cutout image and mask are generated in the same way as mentioned above. However, for {\it Spitzer} and {\it WISE} images, the masking of contaminants is unsatisfactory because these instruments' PSF sizes are much larger and have bright and extended wings. Therefore, other sources on the images are fitted with PSF models at the same time as the target dwarf galaxy (cf.\ \citealt{Janssens2022}). These contamination sources are found by SExtractor and supplemented by eye, with magnitudes as faint as $\sim 21$ mag in {\it WISE} W1 and W2 bands and about $\sim 23$ mag in the {\it Spitzer} $3.6\mu$m band. Although a small fraction of these sources are background galaxies, they are generally unresolved by {\it Spitzer} and {\it WISE} and can be treated as point sources. This approach to contaminants differs from the initial PSF-fitting and masking used by \citet{Buzzo2022}. The new approach makes little difference for the relatively bright dwarfs in common (DF2 and DF4), but is critical for the fainter dwarfs in our sample.

None of the galaxies in our sample are detected in either of the mid-IR bands of {\it WISE}, W3 and W4. To estimate the upper limit of flux in these two bands, we first generate elliptical apertures corresponding to $R_{\mathrm e}$ and $b/a$ for each galaxy. Next, we position these apertures at random locations and position angles on the image, and measure the flux within the apertures -- assuming in effect that the galaxies exhibit the same $R_{\mathrm e}$ and $b/a$ in both mid-IR and optical bands. We tested brighter trail dwarfs by setting their morphological parameters of {\it WISE} and {\it Spitzer} images to be free in GALFIT, and found that their $R_{\rm e}$ values are consistent with optical imaging results. We set the magnitude limit to correspond to twice the standard deviation of the flux measured in these randomly placed apertures.

GALFIT provides unrealistically small fitting uncertainties, so we recalculate the uncertainties using Monte Carlo simulations. First, we apply Gaussian perturbations to each image based on the error map. As a rough correction for the correlated noise between pixels, we magnify the error map by a factor of $\sqrt{\rm NEA}$, where the Noise Equivalent Area ${\rm NEA}=4\pi\times({\rm FWHM_{PSF,pixels}}/2.355)^2$. Then, for each set of perturbations for all images, we repeat the fitting procedure with GALFIT as described above. We repeat this process 500 times and obtain a suite of measurements of structural parameters, magnitudes and colors. We take the standard deviation of each parameter as an estimate of measurement uncertainty (while any additional errors from sky background and photometric calibration are usually much smaller than this Monte Carlo uncertainty). However, we note that the S\'ersic model choice and the details of the masking could give systematic uncertainties of $\sim 0.03$ mag, which we do not include in our error budget. We find consistent magnitudes within the uncertainties for different bands with similar wavelengths. We show the GALFIT best-fitting models for all our galaxies in Figure \ref{fig:galfit_all} in Appendix~\ref{morph_appendix}.

Tables \ref{tab:data_properties} and \ref{tab:data_photometry} summarize the properties and photometric results for each galaxy in our sample. In our sample, we typically obtain the deepest images from {\it HST}. DECaLS and CFHT provide images of comparable quality and depth. For the infrared data, we find good agreement in magnitudes between {\it Spitzer} 3.6 $\mu$m and {\it WISE} W1, but the uncertainties in magnitudes from {\it WISE} are considerably larger than those from {\it Spitzer}. The infrared photometry is not performed for RCP~8, RCP~14, RCP~17, and RCP~32 because their SB levels are too low for meaningful measurements.

Our photometry and colors exhibit reasonable consistency compared to measurements from the literature of galaxies in common, despite various differences in the methods used (e.g., \citealt{Cohen2018,Roman2021,vanDokkum2022b,Buzzo2022}). In particular, \cite{Roman2021} have the most sample overlap with us, and we do not find any significant differences from their results.

\subsection{Photometric Recalibration for CFHT and {\it HST}}\label{sec:recal}
\label{zpcorr}

Prior to combining data from different telescopes for SED fitting, it is essential to ensure that the flux calibration between them is in good agreement. For DECaLS and CFHT data, we select bright stars ($g <20$ mag) within the field of view to compare with the Sloan Digital Sky Survey (SDSS). We find no significant magnitude differences between DECaLS and SDSS. However, the $u$, $g$ and $i$ bands of CFHT are found to be $0.096 \pm 0.012$, $0.118 \pm 0.007$, and $0.091 \pm 0.005$ mag brighter, respectively, than the magnitudes predicted from the SDSS photometric system\footnote{\href{https://www.cadc-ccda.hia-iha.nrc-cnrc.gc.ca/en/megapipe/docs/filt.html}{https://www.cadc-ccda.hia-iha.nrc-cnrc.gc.ca/en/megapipe/docs/filt.html}}. There is no substantial variation in offset between different CCDs.

For the {\it HST} data, we encounter a similar flux calibration disagreement for unknown reasons. We realize that the {\it HST} zeropoints are extremely well determined, but our goal is to bring them to the same system as the rest of our photometry. The systematic error probably comes from the ground-based filters. However, we lack a sufficient number of bright stars in common between {\it HST} and SDSS or DECaLS to perform the same recalibration as described above. Instead, we use the DECaLS photometry to predict galaxy magnitudes in $V_{606}$ and $I_{814}$ for {\it HST}. We run {\tt PROSPECTOR} SED fitting (see Section \ref{prospector}) with only the $griz$ bands from DECaLS for the dwarf galaxies in our sample with {\it HST} observations, and obtain the best-fitting model magnitudes in the two {\it HST} bands as the prediction. We find a relatively constant offset where the measured {\it HST} magnitudes are fainter than the predicted values: $0.066 \pm 0.005$ for the $V_{606}$ band and $0.066 \pm 0.005$ for the $I_{814}$ band, respectively.

Considering that these offsets are relatively constant and do not vary significantly with the brightness of the source, we regard this as a zero-point issue. As a result, we apply corrections to the magnitudes measured from CFHT and {\it HST} data, and incorporate the scatter of the offsets as an additional uncertainty term in the photometry.

\subsection{{\tt PROSPECTOR}}\label{sec:prosp}
\label{prospector}

\begin{figure*}
\centering
\vspace{-5pt}
\includegraphics[width=1\textwidth]{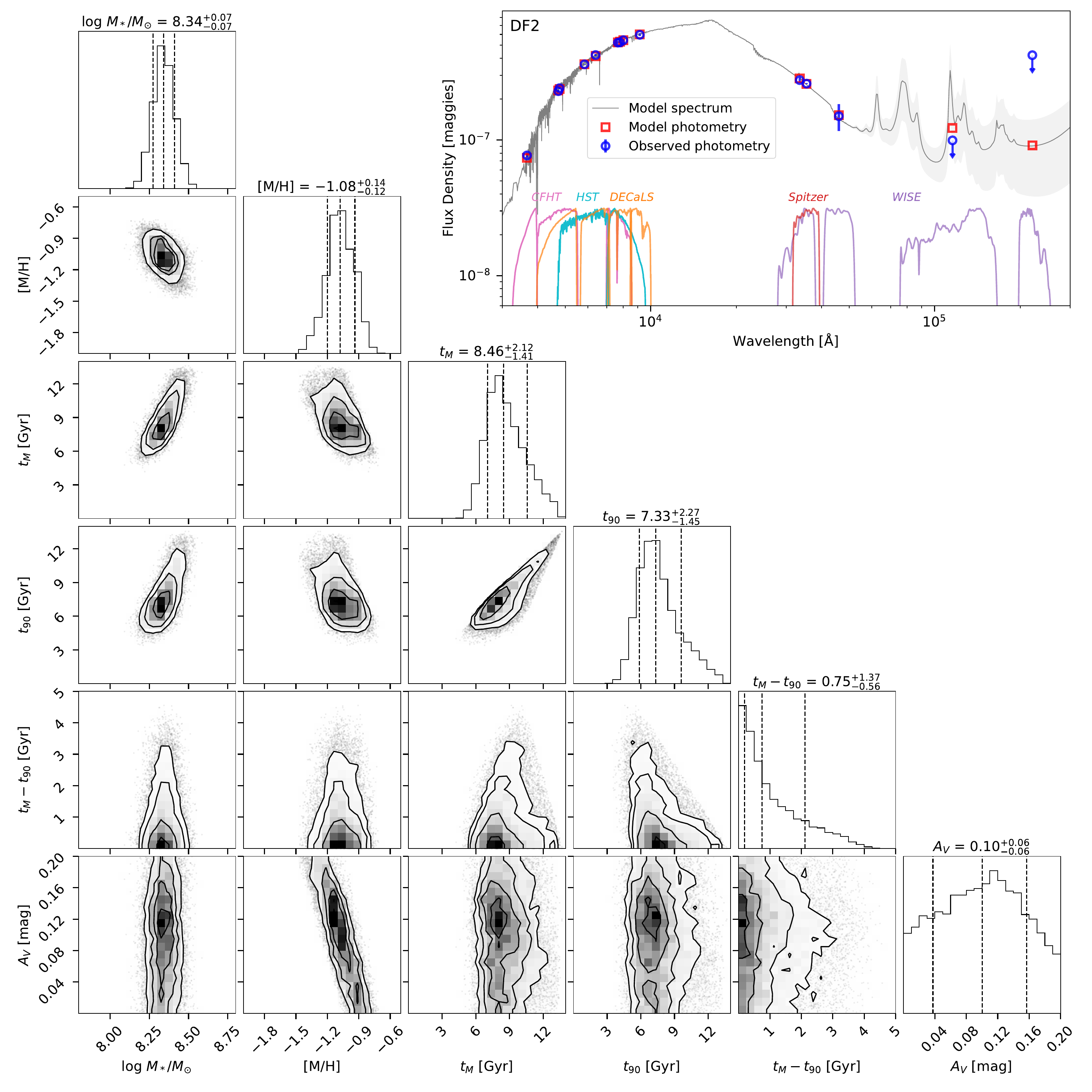}
\caption{DF2 as an example of stellar population inference from $u$-band through mid-IR photometric SED fitting with {\tt PROSPECTOR}. The upper right panel shows the observed SED (blue circles), model SED (red squares), and model spectrum (gray curve with shaded region showing model uncertainty). The upper limits of {\it WISE} W3 and W4 are 1-$\sigma$ limits. At the bottom of this panel, we plot all the filters used in the fitting, which differ slightly from galaxy to galaxy (DF2 is an example with the most complete filter set). The lower left panels show one-dimensional (histograms) and two-dimensional (contours and shaded regions) projections of the posterior probability distribution function (PDF) for five parameters of the stellar populations. On the top of each histogram is the median value of the posterior for that corresponding parameter, with the errorbars giving the 16th and 84th percentiles. The stellar population parameters are well constrained except for $t_M-t_{90}$. Some modest degeneracies can still be found among age, metallicity and dust. The redshift posterior is not shown here because of the poor constraint. The complete figure set for all our dwarfs (25 images) is available in the online journal, except RCP~8, RCP~14, RCP~17 and RCP~32.}
\label{fig:prospector_example}
\end{figure*}

We carry out our SED fitting for all the galaxies in our sample with the Bayesian inference code \texttt{PROSPECTOR} (1.4.0; \citealt{Johnson2021}). Our stellar components are modeled with the MIST isochrones \citep{Choi2016} by the Flexible Stellar Population Synthesis code ({\tt FSPS} 3.2; \citealt{Conroy2009}), based on the MILES stellar spectral library \citep{Falcon-Barroso2011}. We adopt the \cite{Kroupa2001} initial mass function. We choose the Small Magellanic Cloud bar extinction curve \citep{Gordon2003} as our dust attenuation model because it has been suggested as a preferred model for dwarf galaxies \citep{Salim2020}. We note that the fitting results are very similar when using alternative dust model options (e.g.,\ the \citealt{Calzetti2000} extinction curve). Dust emission in the infrared is also included in our fitting.

A parametric star formation history is assumed, wherein the star formation rate (SFR) decreases exponentially over time and is then truncated at some point, as shown by the function below:
\begin{equation}
{\rm SFR}(t) \propto
\begin{cases}
e^{-(t_{\rm start}-t)/\tau} & \text{, } t_{\rm trunc} < t < t_{\rm start} \\
0 & \text{, } 0 < t < t_{\rm trunc}
\end{cases}
\end{equation}
where $t_{\rm start}$ and $t_{\rm trunc}$\footnote{These two parameters are called \texttt{tage} and \texttt{sf\_trunc} in \texttt{PROSPECTOR}.} mean the lookback time when star formation starts and truncates separately, and $\tau$ is the e-folding timescale. This relatively flexible model should provide a reasonable and physically-motivated approximation to the SFHs of the galaxies in our sample. Most of our sample are LSB dwarfs, which \cite{Greco2018} found previously could be well modeled with exponentially declining SFHs. The extra parameter $t_{\rm trunc}$ allows (but does not require) the modeling of a star formation shutdown caused by strong stellar feedback in the bullet dwarf scenario or by gas stripping during dwarf galaxy infall into a group.

We set linearly uniform priors on all of our seven free parameters in the SED fitting, including redshift ($0.003 < z < 0.007$, which brackets a velocity range of $\pm 600$~km~s$^{-1}$ relative to NGC 1052), log stellar mass ($6 < \log (M_\star/\mathrm{M}_{\odot}) < 9$), metallicity ($-2.0 < {\rm [M/H]} < -0.5$), star formation start time ($1.0 < t_{\rm start} < 13.8$ Gyr), star formation truncation time ($0 < t_{\rm trunc}/t_{\rm start} < 1$), log e-folding timescale ($0.01 < \tau < 20$ Gyr) and dust extinction ($0 < A_V < 0.2$ mag, since quiescent dwarf galaxies are generally thought to be dust-free). Here we note that small amounts of dust ($A_V \sim$~0.1 mag) in these results can be considered as a fictitious component that helps the fitting to compensate for unidentified systematic problems with the stellar populations synthesis models or with the photometry (see discussion in Appendix~\ref{stellar_pop_appendix}). Nevertheless, our results and conclusions are essentially unchanged if we run the fitting without dust.

We use the dynamic nested sampling algorithm \texttt{dynesty} \citep{Speagle2020} to sample the posteriors. Compared to the {\tt PROSPECTOR} settings in \cite{Buzzo2022}, our priors and star formation history model have only minor changes to reflect the NGC~1052 group and the potential formation mechanisms of trail dwarfs. We note that our results still hold if we use a similar configuration to that of \cite{Buzzo2022}.

Although our parametric SFH model is based on three parameters $t_{\rm start}$, $t_{\rm trunc}$ and $\tau$, we note that they are not the most robust parameters we can recover from the SED fitting. We also calculate two complementary parameters from the SFH curves: mass-weighted age $t_M$, and $t_{90}$, the lookback time when 90\% of the total stellar mass has been formed (similar to the quenching time). We will report these more robust quantities in our analysis, along with $t_M-t_{90}$ as an indicator of star formation timescale (which is numerically close to $t_{50}-t_{90}$). The results of our SED fitting are given in Table \ref{tab:data_sedfitting}, with an example fit shown in Figure~\ref{fig:prospector_example}. The fitting results of all other galaxies are shown in an online-available figure set attached with Figure \ref{fig:prospector_example}.

Although some degeneracy among stellar mass, age and metallicity can still be seen in the corner plots of the fitting results, we get excellent fits and acceptable constraints on the stellar population parameters for most galaxies. By running {\tt PROSPECTOR} without some of the filters, we find that the $u$-band from CFHT is helpful in constraining the ages, particularly for the fainter galaxies. Near-IR data effectively determine metallicity, even for brighter galaxies like DF2 and DF4. However, among all parameters, the star formation timescale $t_M-t_{90}$ almost always has large uncertainties and is sensitive to priors on $\tau$ and $t_{\rm trunc}/t_{\rm start}$. We also tested different parameterized forms of the star formation history, such as the tau model without truncation and the delayed tau model. These results remained consistent within the uncertainties with those from our fiducial star formation history. 

Spectroscopy is generally considered to provide the gold standard for stellar population constraints in galaxies (relative to SED fitting), and the availability of spectra for three galaxies in our sample allows us to cross-check the two methods. Details are presented in Appendix~\ref{stellar_pop_appendix}, with the finding that the SED-fitting and spectroscopy results are reasonably consistent, while the spectroscopic constraints with limited wavelength range are remarkably sensitive to continuum corrections. Additional tests of the SED fitting results are also discussed in Appendix~\ref{stellar_pop_appendix}.

\section{Morphologies}
\label{results_morph}

\begin{figure}
\centering
\includegraphics[width=0.475\textwidth]{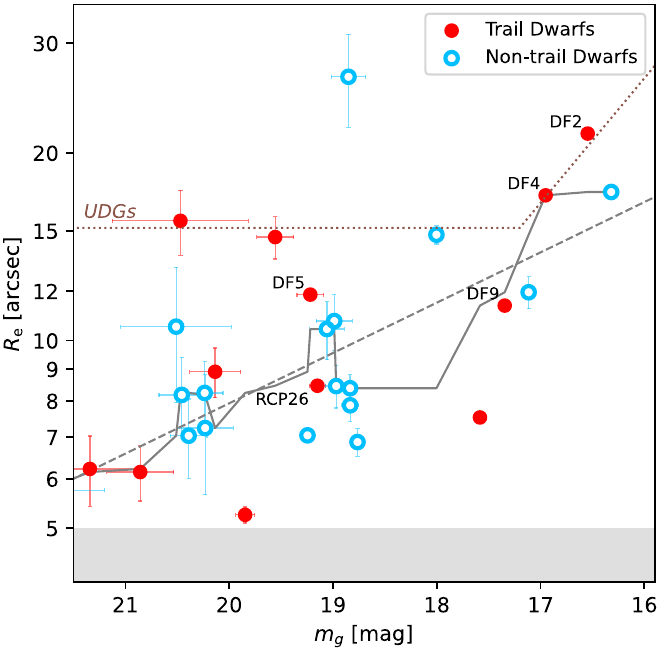}
\caption{The distribution of semi-major axis effective radius and $g$-band magnitude of trail dwarfs (red solid points) and non-trail dwarfs (blue open points). The gray band shows that our sample excludes galaxies with $R_{\rm e}<5$~arcsec as possible background sources. The brown dotted line shows the lower boundary of the UDG region (the diagonal segment is an approximate SB limit, since circularized $R_{\rm e}$ is not plotted here). The gray solid line is a running median with $N=7$, and the gray dashed line is a least-squares fit to the running median. After subtracting this fit, we find no significant size difference between trail and non-trail dwarfs.}
\label{fig:mag_reff}
\end{figure}

\begin{figure*}
\centering
\includegraphics[width=0.73\textwidth]{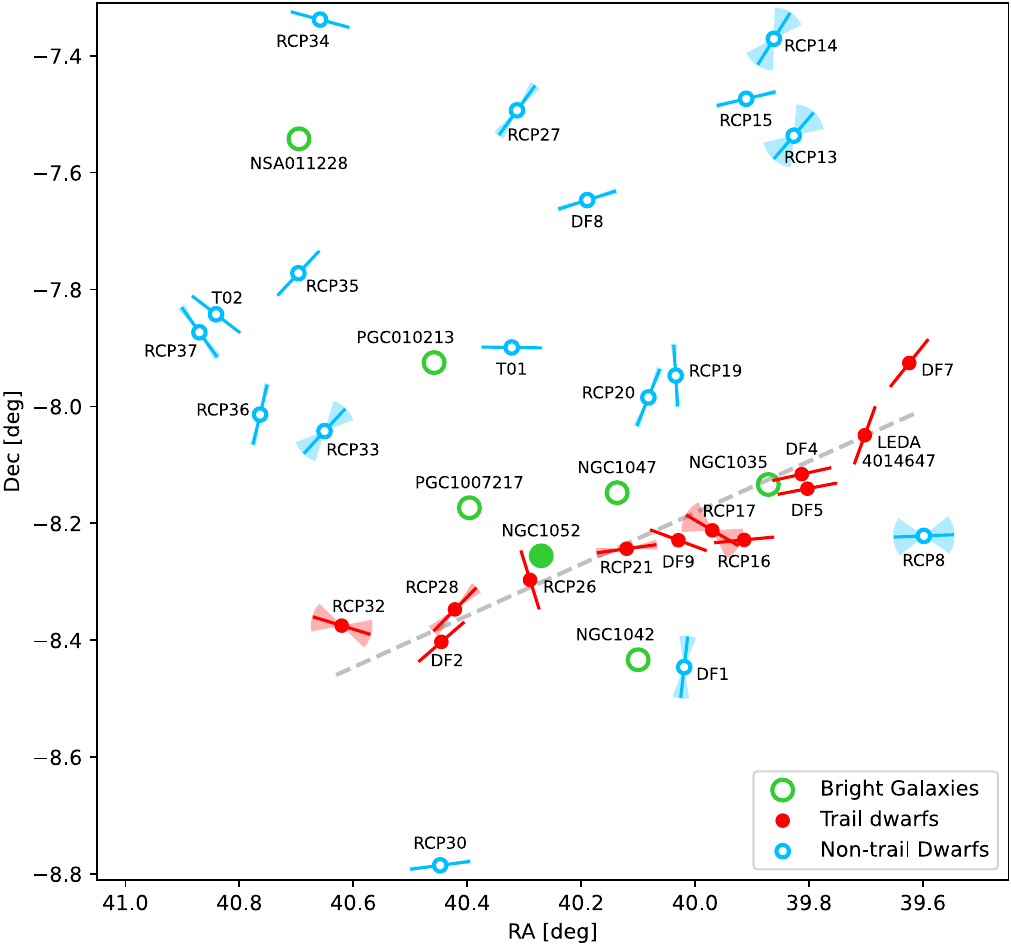}
\caption{Sky positions of trail dwarfs (red solid points) and non-trail dwarfs (blue open points), with isophote position angles (PAs) represented by the direction of the line segment through each galaxy. Galaxies without line segments were too faint to measure the PA. The shading around the line segments represents the PA measurement uncertainty. The massive central galaxy NGC~1052 is shown by a solid green circle, and other non-dwarf galaxies in this sky region as open green circles. Most of the trail dwarfs align parallel to the trail itself (marked with a gray dashed line from \citealt{vanDokkum2022a}), with RCP~26 as an exception. The non-trail dwarfs are also preferentially aligned in the same direction, although with a lower significance.}
\label{fig:pa}
\end{figure*}

Analysis results are reported in the next three Sections, followed by theoretical interpretation in Section~\ref{discussion_theories}. In general, we are testing for characteristics of the trail dwarfs that are distinct from those of non-trail dwarfs, to help diagnose differences in formation histories, and to provide independent tests for the existence of the trail as a physically meaningful structure. Here it should be kept in mind that the trail sample is likely to include some contaminant galaxies whose sky positions happen to overlap.

Visual inspection of the optical imaging shows most of the trail dwarfs to have smooth and symmetric morphologies, like relaxed, quiescent spheroidal systems. The exceptions are DF7, which appears very elongated; RCP~26, with suggestions of outer isophote distortions; and the lowest SB objects whose shapes are indistinct (RCP~17, RCP~21, RCP~28, RCP~32). The non-trail dwarfs appear qualitatively similar, with the exceptions of DF8 and RCP~34 which show asymmetries and blue patches of recent star formation; RCP~36 may also be asymmetric. Two trail dwarfs (DF9 and LEDA 4014647) and one non-trail dwarf (T01) seem to have nuclear star clusters, which in the case of DF9 has previously been studied in detail \citep{Gannon2023}.

Moving to quantitative metrics of morphology, we first re-examine the findings of \citet[figure~4]{vanDokkum2022a} that trail dwarfs have unusually large sizes at fixed magnitude, on average, or equivalently have relatively low SB. We plot the size--magnitude distribution in Figure \ref{fig:mag_reff}. To remove the magnitude dependency of size, we calculate a running median of $R_{\rm e}$ with $N=7$ and then conduct a least-squares fit to it, following the same procedure as \cite{vanDokkum2022a}. We note that $N=7$ is a good choice here to avoid noise when $N$ is too small and oversmoothing when $N$ is large. The best-fit line has the form $\log R_{\rm e} / {\rm kpc} =-0.08\times (g-20)+0.90$, very close to the result for a sample including dwarfs beyond the virial radius in \cite{vanDokkum2022a}. After subtracting this fit, we apply the two-sample Wilcoxon rank-sum test to the sizes without magnitude dependency between trail and non-trail dwarfs. The median size of trail dwarfs is 7\% larger than non-trail dwarfs, but the difference is not significant ($p$-value of 0.6). This result does not qualitatively change if we make the size comparison in the redder filters which trace the stellar mass better, or directly use stellar mass. The possible variation in line-of-sight distance does not bring a significant change to the distributions. Otherwise, our different conclusion from \cite{vanDokkum2022a} is mainly caused by a different sample selection, as we have restricted the control sample of non-trail dwarfs to come from the same group-centric distances as the trail dwarfs. In contrast, \citet{vanDokkum2022a} included dwarfs that are much farther out and tend to have higher SB, perhaps owing to younger ages. We note that a smaller sample size could decrease the significance as well. We have tested their analysis after splitting their data set into two groups of galaxies inside and outside the virial radius, and found a significant size difference between them, with a $p$-value of 0.01 in the Wilcoxon rank-sum test.

DF2 was originally identified as a galaxy of interest because it is an ultra-diffuse galaxy (UDG) with a populous GC system. Using a definition of UDGs based on \citet{vanDokkum2015}, $R_{\rm e} \geq 1.5$~kpc and $\langle \mu_g \rangle_{\rm e} \gtrsim 25.1$, we find six UDGs in our sample (DF1, DF2, DF4, DF7, RCP~32, T02), with four of them borderline cases (see Figure~\ref{fig:mag_reff}). RCP~34 is formally too bright, but would fade into the UDG region if its star formation were to cease. An empirical scaling relation for number of UDGs versus halo mass \citep{Goto2023} predicts three UDGs for the NGC~1052 group ($6\times10^{12} M_\odot$ virial mass; \citealt{Forbes2019}), and an observation of six UDGs is within the uncertainty and scatter. We also note that UDGs with extremely low SB, such as RCP~32, may be missed in previous surveys, so this empirical relation could underestimate the UDG numbers.

We next examine the position angles (PAs) of the semimajor axes of the dwarfs, with the GALFIT results corresponding approximately to the PA at $R_{\rm e}$. Figure \ref{fig:pa} shows the sky positions of trail and non-trail dwarf galaxies, with PAs indicated by line segments. The PAs of the trail galaxies appear remarkably well aligned with the direction of the trail. There is also a visual suggestion that the trail PAs are aligned locally with an S-shaped curvature in the trail. RCP~26 is an exception, with a PA orthogonal to the trail, and we have already noted that this galaxy shows signs of peculiar isophote twists. 

A PA alignment among the trail dwarfs initially appears to be new evidence in support of physical association between these objects. However, we notice that about eight non-trail dwarfs also have their PAs aligned with the direction of the trail. If we define alignment as a $\lesssim 30^\circ$ difference (which is nearly equal to the PA standard deviation of the trail dwarfs) between galaxy PA and trail PA ($=-66^\circ$), then 8 out of 17 non-trail dwarfs are aligned with a significance level of 0.17 compared to random orientations. The trail dwarf alignment (8 out of 12 dwarfs) is much less likely to be random, with a significance level of 0.02 (and if we consider a ``look-elsewhere'' effect where an orthogonal alignment would have also been noteworthy, the result is still significant). 

We also carry out a ``blind'' test for PA clustering, using the Rayleigh test which is designed to identify a non-uniform distribution around a circle. We find non-uniformity with a significance level of $p=0.14$ and 0.22, respectively, for the trail and non-trail dwarfs (if RCP~26 were removed from the trail sample because it appears to be tidally interacting with NGC~1052, the alignment has $p=0.04$ significance). This test provides mild evidence for alignment among the trail dwarfs, with a much higher significance level found when folding in the prior information about the PA of the trail. We furthermore use the two-sample Kuiper test to compare the PA distribution between the trail and non-trail dwarfs. The Kuiper test is similar to the Kolmogorov–Smirnov test, but has invariance under cyclic transformations, making it more suitable for analyzing PAs. We get a $p$-value of 0.6, which suggests no significant PA distribution difference between them. A group-wide alignment of PAs is puzzling to explain, unless it traces an alignment of galaxy angular momenta with the direction of filamentary infall (e.g.,\ \citealt{Rong2020}), which may be perpendicular to the trail based on the large-scale distribution of galaxies in the NGC~1052 region \citep{vanDokkum2022a}.

\begin{figure*}
\centering
\includegraphics[width=0.99\textwidth]{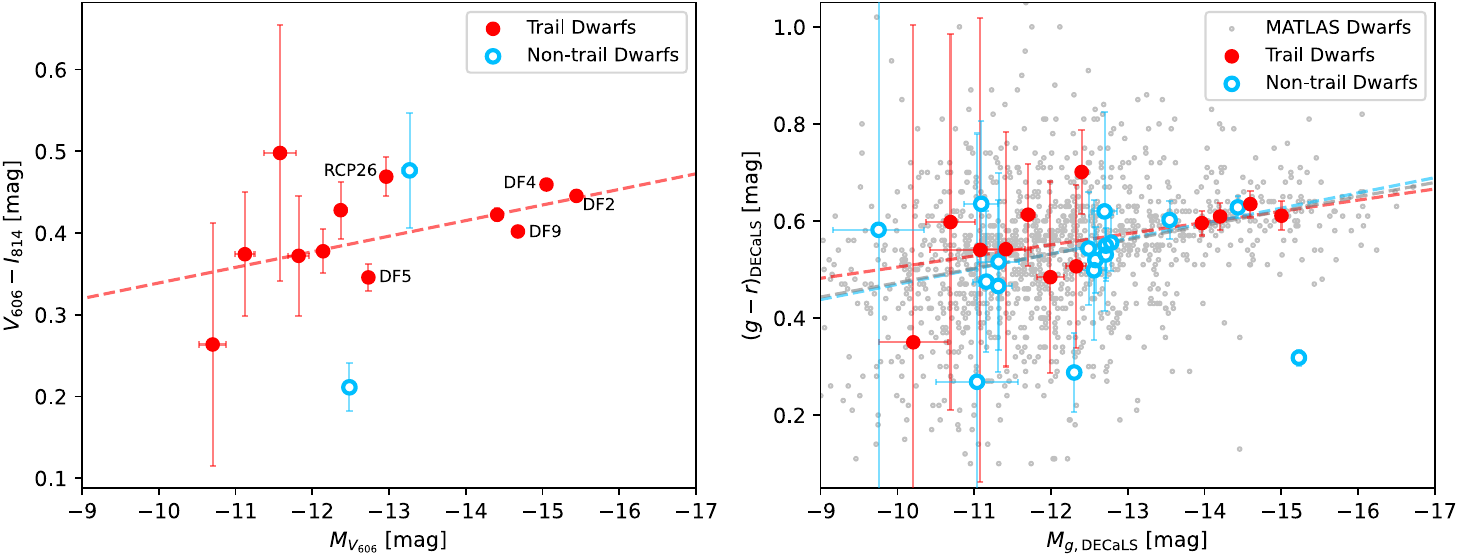}
\caption{Color--magnitude relations (CMRs) of trail dwarfs (red solid points), non-trail dwarfs (blue open points) and MATLAS dwarfs (gray smaller points; \citealt{Poulain2021}). The dashed lines are the best linear fits for each galaxy sample. The two panels show $V_{606}-I_{814}$ color from {\it HST} and $g-r$ color from DECaLS. The photometry is corrected for Galactic extinction. The original CFHT $g-r$ colors of MATLAS dwarfs are transformed here to DECaLS colors, shown in the right panel. Almost all the trail dwarfs seem to fall along a tight red sequence in both panels, except for RCP~26 with redder colors. Most of the non-trail dwarf galaxies, and the MATLAS dwarfs, are consistent with following the same red sequence as the trail dwarfs. Either these galaxy samples share similar stellar populations, or there are degeneracies that preclude detecting differences with CMRs alone.}
\label{fig:cmd}
\end{figure*}

A summary of all the morphological parameters obtained by GALFIT is provided in Figure~\ref{fig:morphology_all}, as a corner plot and histograms. There is no significant difference in the S\'ersic index between trail and non-trail dwarfs: median $n = 0.65 \pm 0.04$ and $0.76 \pm 0.04$, respectively (with uncertainties estimated through bootstrapping). There is a hint that the trail dwarfs are typically rounder, but the result is not significant given the sample sizes: $b/a = 0.82 \pm 0.03$ versus $0.74 \pm 0.07$. Also, any difference might be driven by a systematic dependence of flattening on groupcentric radius due to tidal effects \citep{Lokas2012,Lokas2015,Barber2015}, since the trail dwarfs are located closer to the center (in projection). As an additional comparison, the MATLAS Survey \citep{Habas2020} provides a large sample of dwarf galaxies in galaxy groups hosted by massive early-type galaxies. The NGC~1052 group was not in the sky region covered by MATLAS, but has a similar environment, allowing for a fair comparison with MATLAS data. The MATLAS dwarfs have a median $b/a = 0.74$ \citep{Poulain2021}, reinforcing the suggestion that the trail dwarfs may be unusually round.

Overall, we find no strong evidence that trail dwarfs are morphologically distinct. There are hints that they have preferentially round shapes and PA alignments with the trail. These properties could be investigated further by increasing the sample size through deeper imaging of the faintest dwarfs. Also, a fairer comparison could be conducted based on purer samples of trail and non-trail dwarfs, if additional diagnostic information (e.g., distance or velocity) is obtained in the future besides sky positions. The trail dwarfs that are potential morphological outliers from the rest are DF7 and RCP~16 (more flattened) and RCP~26 (different PA). If the trail structure is actually very extended along the line of sight (i.e., following the $\sim$~2~Mpc line-of-sight separation between DF2 and DF4; \citealt{Shen2021b,Shen2023}), then the round shapes could be caused by the extreme foreshortening of the geometry, with the trail dwarfs having greatly elongated, prolate intrinsic shapes.

We note that the morphology measurements of the trail dwarfs all come from {\it HST} imaging, but from shallower DECaLS data for most of the non-trail dwarfs. However, we do not find qualitative changes to our results if we obtain the morphological parameters of the trail dwarfs using DECaLS imaging instead (Appendix~\ref{morph_appendix} discusses the consistency between the two datasets).

\section{Color--magnitude relations}
\label{results_cmr}

\begin{deluxetable*}{lcccccc}
\tabletypesize{\footnotesize}
\renewcommand{\arraystretch}{1.3}
\tablewidth{0pt}
\tablecaption{Results from {\tt PROSPECTOR} SED fitting of the dwarf galaxies in our sample, sorted by right ascension. From left to right, the columns show galaxy names, stellar mass $\log M_*/M_{\odot}$, stellar metallicity [M/H], mass-weighted stellar age $t_M$, look-back time when a galaxy has formed 90\% of its stellar mass $t_{90}$, star-formation timescale $t_M-t_{90}$, and $g$-band mass-to-light ratio $M_*/L_g$.
\label{tab:data_sedfitting}}
\tablehead{
\colhead{Galaxy} & \colhead{$\log M_*/M_{\odot}$} & \colhead{[M/H]} & \colhead{$t_M$} & \colhead{$t_{90}$} & \colhead{$t_M-t_{90}$} & \colhead{$M_*/L_g$} \\
\colhead{} & \colhead{} & \colhead{[dex]} & \colhead{[Gyr]} & \colhead{[Gyr]} & \colhead{[Gyr]} & \colhead{[$M_{\odot}/L_{\odot,g}$]}}
\startdata
    DF7 (RCP 9) & $7.02_{-0.19}^{+0.14}$ & $-1.21_{-0.47}^{+0.44}$ & $6.73_{-3.30}^{+3.74}$ & $5.45_{-2.96}^{+4.06}$ & $0.68_{-0.49}^{+1.30}$ & $1.55_{-0.55}^{+0.57}$\\
    LEDA 4014647 & $7.98_{-0.06}^{+0.06}$ & $-1.38_{-0.14}^{+0.11}$ & $11.54_{-1.69}^{+1.12}$ & $10.83_{-1.99}^{+1.42}$ & $0.49_{-0.33}^{+0.79}$ & $2.32_{-0.24}^{+0.19}$\\
    DF5 (RCP 11) & $7.15_{-0.11}^{+0.10}$ & $-1.76_{-0.16}^{+0.25}$ & $7.49_{-2.39}^{+2.90}$ & $6.40_{-2.34}^{+3.01}$ & $0.70_{-0.51}^{+1.28}$ & $1.54_{-0.32}^{+0.38}$\\
    DF4 (RCP 12) & $8.25_{-0.06}^{+0.06}$ & $-1.04_{-0.12}^{+0.07}$ & $10.68_{-1.51}^{+1.38}$ & $9.90_{-1.64}^{+1.61}$ & $0.55_{-0.39}^{+0.84}$ & $2.43_{-0.25}^{+0.18}$\\
    RCP 13 & $7.21_{-0.21}^{+0.14}$ & $-1.29_{-0.45}^{+0.45}$ & $6.75_{-3.42}^{+3.70}$ & $5.51_{-3.11}^{+4.03}$ & $0.65_{-0.47}^{+1.27}$ & $1.56_{-0.58}^{+0.57}$\\
    RCP 15 & $6.58_{-0.23}^{+0.19}$ & $-1.56_{-0.31}^{+0.47}$ & $4.94_{-2.70}^{+4.05}$ & $3.74_{-2.21}^{+4.08}$ & $0.63_{-0.45}^{+1.26}$ & $1.06_{-0.43}^{+0.58}$\\
    RCP 16 (Ta21-12000) & $6.97_{-0.13}^{+0.10}$ & $-1.38_{-0.35}^{+0.33}$ & $8.16_{-3.02}^{+2.96}$ & $7.01_{-2.93}^{+3.27}$ & $0.70_{-0.51}^{+1.22}$ & $1.81_{-0.44}^{+0.41}$\\
    DF1 (RCP 18) & $7.40_{-0.15}^{+0.11}$ & $-1.36_{-0.34}^{+0.41}$ & $7.83_{-3.18}^{+3.06}$ & $6.62_{-3.14}^{+3.41}$ & $0.71_{-0.51}^{+1.35}$ & $1.96_{-0.55}^{+0.51}$\\
    DF9 & $8.05_{-0.07}^{+0.06}$ & $-1.24_{-0.12}^{+0.11}$ & $10.36_{-1.95}^{+1.61}$ & $9.52_{-2.17}^{+1.86}$ & $0.60_{-0.43}^{+0.95}$ & $2.19_{-0.25}^{+0.22}$\\
    RCP 19 & $7.24_{-0.16}^{+0.12}$ & $-1.53_{-0.31}^{+0.37}$ & $7.13_{-3.05}^{+3.28}$ & $5.88_{-2.81}^{+3.56}$ & $0.73_{-0.53}^{+1.29}$ & $1.54_{-0.46}^{+0.44}$\\
    RCP 20 & $7.36_{-0.14}^{+0.11}$ & $-1.50_{-0.31}^{+0.37}$ & $7.62_{-2.94}^{+3.21}$ & $6.43_{-2.86}^{+3.49}$ & $0.72_{-0.53}^{+1.29}$ & $1.63_{-0.44}^{+0.43}$\\
    RCP 21 & $6.78_{-0.21}^{+0.14}$ & $-1.29_{-0.44}^{+0.46}$ & $6.48_{-3.33}^{+3.65}$ & $5.20_{-2.92}^{+3.96}$ & $0.67_{-0.48}^{+1.29}$ & $1.53_{-0.58}^{+0.57}$\\
    DF8 (RCP 24) & $6.63_{-0.12}^{+0.13}$ & $-1.72_{-0.20}^{+0.37}$ & $1.78_{-0.65}^{+1.01}$ & $1.19_{-0.51}^{+0.87}$ & $0.43_{-0.27}^{+0.63}$ & $0.48_{-0.12}^{+0.15}$\\
    RCP 26 & $7.40_{-0.09}^{+0.07}$ & $-0.86_{-0.17}^{+0.17}$ & $9.43_{-2.55}^{+2.18}$ & $8.37_{-2.65}^{+2.56}$ & $0.67_{-0.49}^{+1.18}$ & $2.57_{-0.45}^{+0.37}$\\
    RCP 27 & $6.65_{-0.25}^{+0.18}$ & $-1.41_{-0.41}^{+0.50}$ & $5.81_{-3.25}^{+4.09}$ & $4.55_{-2.78}^{+4.34}$ & $0.62_{-0.44}^{+1.26}$ & $1.25_{-0.53}^{+0.61}$\\
    T01 & $8.10_{-0.13}^{+0.11}$ & $-0.77_{-0.20}^{+0.17}$ & $7.09_{-2.62}^{+3.22}$ & $5.80_{-2.39}^{+3.53}$ & $0.73_{-0.54}^{+1.37}$ & $1.96_{-0.49}^{+0.54}$\\
    RCP 28 & $6.49_{-0.20}^{+0.14}$ & $-1.38_{-0.41}^{+0.47}$ & $6.61_{-3.36}^{+3.61}$ & $5.27_{-2.94}^{+3.97}$ & $0.67_{-0.49}^{+1.32}$ & $1.55_{-0.58}^{+0.56}$\\
    DF2 (RCP 29) & $8.34_{-0.07}^{+0.07}$ & $-1.08_{-0.14}^{+0.15}$ & $8.28_{-1.41}^{+2.07}$ & $7.12_{-1.42}^{+2.17}$ & $0.80_{-0.60}^{+1.38}$ & $1.99_{-0.20}^{+0.28}$\\
    RCP 30 & $7.31_{-0.15}^{+0.12}$ & $-1.49_{-0.33}^{+0.39}$ & $7.27_{-3.07}^{+3.28}$ & $6.03_{-2.88}^{+3.61}$ & $0.70_{-0.52}^{+1.31}$ & $1.57_{-0.46}^{+0.45}$\\
    RCP 33 & $7.20_{-0.21}^{+0.15}$ & $-1.44_{-0.37}^{+0.47}$ & $6.29_{-3.21}^{+3.78}$ & $5.03_{-2.83}^{+4.02}$ & $0.67_{-0.49}^{+1.28}$ & $1.40_{-0.52}^{+0.54}$\\
    RCP 34 & $7.86_{-0.11}^{+0.08}$ & $-1.80_{-0.14}^{+0.27}$ & $2.08_{-0.65}^{+0.71}$ & $1.51_{-0.54}^{+0.65}$ & $0.41_{-0.26}^{+0.60}$ & $0.56_{-0.12}^{+0.10}$\\
    RCP 35 & $7.32_{-0.13}^{+0.10}$ & $-1.62_{-0.25}^{+0.32}$ & $7.67_{-2.85}^{+2.88}$ & $6.50_{-2.84}^{+3.19}$ & $0.71_{-0.52}^{+1.28}$ & $1.59_{-0.40}^{+0.39}$\\
    RCP 36 & $6.48_{-0.23}^{+0.22}$ & $-1.52_{-0.33}^{+0.47}$ & $4.12_{-2.27}^{+4.12}$ & $3.02_{-1.79}^{+3.96}$ & $0.61_{-0.43}^{+1.20}$ & $0.98_{-0.41}^{+0.63}$\\
    T02 & $7.71_{-0.13}^{+0.10}$ & $-1.17_{-0.33}^{+0.27}$ & $7.85_{-2.91}^{+3.04}$ & $6.59_{-2.68}^{+3.47}$ & $0.70_{-0.51}^{+1.27}$ & $1.81_{-0.44}^{+0.43}$\\
    RCP 37 & $6.66_{-0.23}^{+0.15}$ & $-1.35_{-0.43}^{+0.49}$ & $6.74_{-3.54}^{+3.64}$ & $5.41_{-3.10}^{+4.03}$ & $0.65_{-0.47}^{+1.30}$ & $1.56_{-0.63}^{+0.63}$\\
\enddata
\end{deluxetable*}

To understand whether trail dwarfs share a similar origin and significantly differ from non-trail dwarfs, we conduct a preliminary exploration of the stellar populations of the two groups of galaxies with color--magnitude relations (CMRs) shown in Figure \ref{fig:cmd} based on different filter combinations. \citet{vanDokkum2022b} found a hint that DF4 is slightly redder than DF2 in the $V_{606}-I_{814}$ color, by $0.02 \pm 0.08$ mag, which we confirm with a color difference of $0.02 \pm 0.01$ mag. We also find that DF4 is consistently redder than DF2 across all bands, which is particularly apparent with the longer wavelength baselines (also found previously by \citealt{Buzzo2022} but with larger uncertainties). The rest of the trail dwarfs are also inconsistent with having uniform colors, and instead show a trend towards bluer colors at fainter magnitudes, like a classical red sequence. RCP~26 is an exception with redder colors, reinforcing the implication from its PA that it is not a true member of the trail. The non-trail dwarf galaxies are on a roughly similar red sequence, except for DF8 and RCP~34, with much bluer colors that, along with somewhat irregular morphologies (Figures~\ref{fig:trail_gallery} and \ref{fig:nontrail_gallery}), suggest recent star formation.

To check if the dwarf galaxies along the trail or in the vicinity are distinct from normal dwarf galaxies, we show a CMR in the right panel of Figure \ref{fig:cmd} that also includes the MATLAS sample. Although ultra-diffuse galaxies (UDGs) represent only a small fraction of the MATLAS Survey, \cite{Marleau2021} found that these UDGs do not show significantly different photometric properties compared to classical dwarf galaxies. It is worth noting that MATLAS uses the CFHT $g$, $r$ and $i$ bands, while our data do not include the CFHT $r$ band, and many galaxies in our sample do not even have CFHT data. Therefore, in Figure \ref{fig:cmd} (right) we plot the CMR of the MATLAS dwarf galaxies using a transformation from CFHT color to DECaLS color. Within the color range of our galaxies of $0.58 \lesssim (g-r)_{\rm DECaLS} \lesssim 0.66$, we apply a mean correction of $g_{\rm DECaLS}-g_{\rm CFHT} \approx 0.046$ and $(g-r)_{\rm DECaLS}-(g-r)_{\rm CFHT} \approx 0.070$, according to {\tt FSPS} single stellar population models (and in agreement with photometry in common for the brighter galaxies in our sample).

We conduct linear fits to the CMRs of each galaxy sample (i.e. trail, non-trail and MATLAS dwarfs), shown in Figure \ref{fig:cmd}. For both trail and non-trail dwarfs, we perform Monte Carlo simulations based on the observed magnitudes and colors and their uncertainties to obtain the linear fitting parameters and errors. For MATLAS dwarfs, since measurement uncertainties are unavailable in the public catalog, we use bootstrapping instead. Outliers that significantly deviate from the linear relation are iteratively rejected in the fitting. We also attempted to include intrinsic scatter as an additional parameter, but the fitting consistently returns scatter $<$~0.005~mag, which is effectively negligible. The fitting result in the {\it HST} diagram is

{\footnotesize $$V_{606}-I_{814} = (-0.019 \pm 0.014) \times (M_{V_{606}}+12.7) + (0.390 \pm 0.024)$$}
and in the $g-r$ panel, we have

{\footnotesize $$g-r = (-0.023 \pm 0.055) \times (M_g+12.7) + (0.567 \pm 0.076) {\rm\ ,trail}$$
$$g-r = (-0.031 \pm 0.060) \times (M_g+12.7) + (0.554 \pm 0.032) {\rm\ ,nontrail}$$
$$g-r = (-0.030 \pm 0.002) \times (M_g+12.7) + (0.552 \pm 0.003) {\rm\ ,MATLAS}$$}
We find no significant differences in the CMRs between the trail dwarfs, non-trail dwarfs and MATLAS dwarf galaxies. We checked the CMRs using other bands ($u-i$ and $g-i$) and found the same trends. In principle, other colors with longer baselines, using DECaLS $z$-band or {\it WISE} W1, should provide more discriminatory power. However, the photometric uncertainties in these two bands are typically too large to provide further useful constraints here. 
 
The red sequence is generally considered as equivalent to a mass--metallicity relation, with scatter contributed by age variations and other processes such as tidal stripping \citep{Collins2022}. Therefore to a first approximation, the trail dwarfs appear to share a similar enrichment history to other dwarfs. On the other hand, degeneracies between age and metallicity can conspire to make different populations appear similar in the CMRs\footnote{We have also attempted to make use of four-dimensional color information using a self-organizing map, including magnitude independent $(g-r)_{\rm DECaLS}$, $(g-i)_{\rm DECaLS}$, $(g-z)_{\rm DECaLS}$ and $g_{\rm DECaLS}-{\rm W1}$ colors which are available for all the galaxies in our sample, and found the trail and non-trail dwarfs have different distributions at the 1-$\sigma$ level.}, and our next step is to disentangle such degeneracies by combining all the bands with SED fitting.

\section{Stellar population results}
\label{results_sed}

\begin{figure}
\centering
\includegraphics[width=0.48\textwidth]{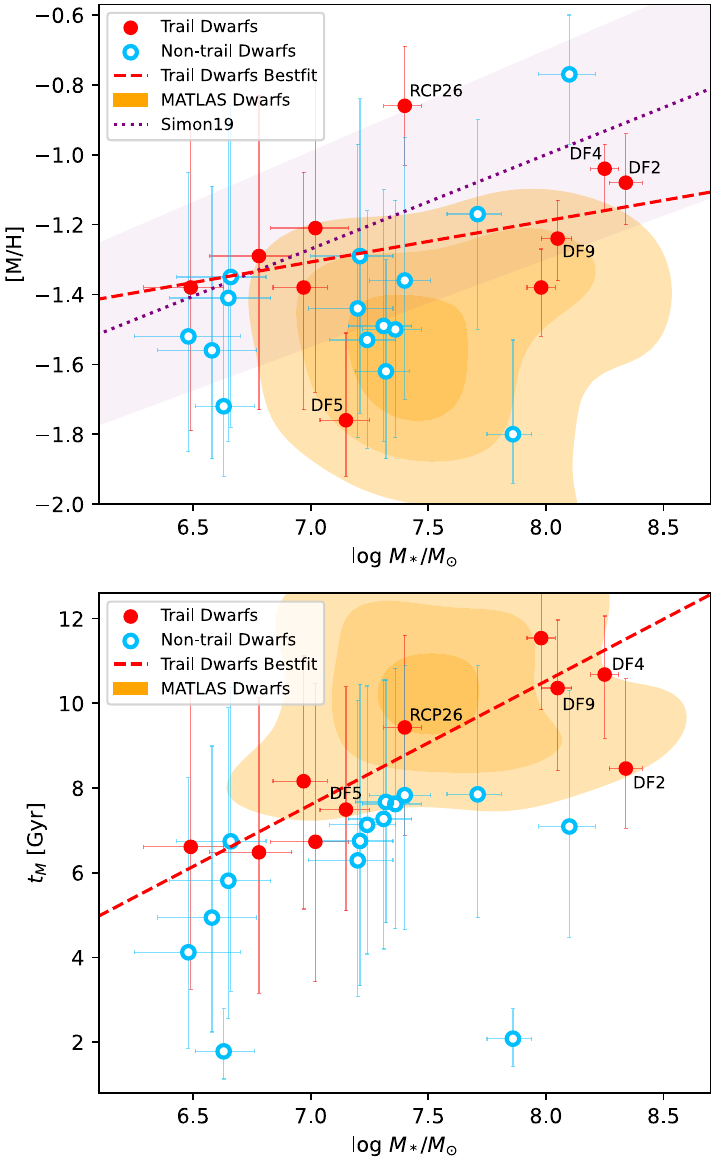}
\vspace{-10pt}
\caption{The distribution of various stellar population properties of trail dwarfs (red solid points), non-trail dwarfs (blue open points) and MATLAS dwarfs (orange contours; \citealt{Heesters2023}). The \cite{Simon2019} Local Group relation is overplotted with purple dotted line and shading. The red dashed line is the best-fit linear relation for trail dwarfs for each panel. The slopes of these relations are likely sensitive to biases in modelling the fainter galaxies, while comparisons between galaxies at similar stellar mass are more robust. {\it Top}: Stellar metallicity vs.\ mass. The NGC~1052 and MATLAS dwarfs have broadly similar mass--metallicity distributions. The trail dwarfs follow a tight relation except for RCP~26 and DF5, which may have different origins compared to other trail galaxies. Non-trail dwarfs are more broadly scattered in the diagram, with most of them lying below the trail-dwarf relation. {\it Bottom}: Mass-weighted stellar age $t_M$ vs.\ stellar mass. The NGC~1052 dwarfs have significantly younger ages than MATLAS dwarfs, and trail dwarfs are older than non-trail dwarfs at fixed mass. Both panels reinforce the hypothesis that trail and non-trail dwarfs are distinct populations.}
\label{fig:mass_metallicity}
\end{figure}

\begin{figure}
\centering
\includegraphics[width=0.48\textwidth]{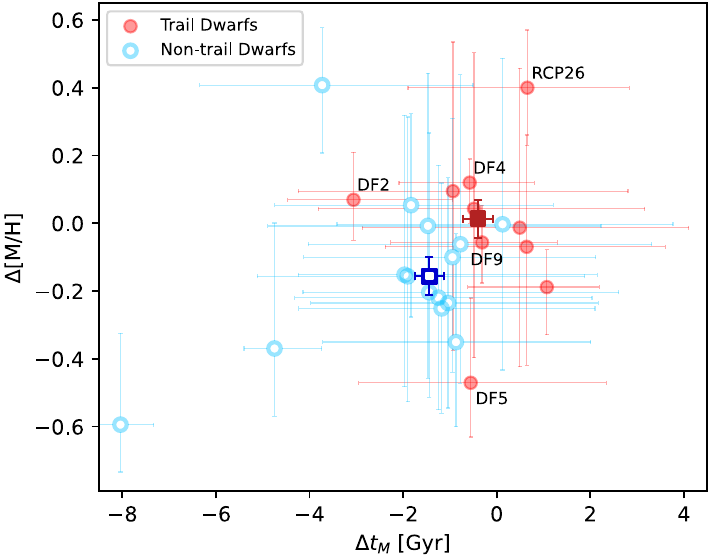}
\vspace{-10pt}
\caption{The distribution of $\Delta$[M/H] and $\Delta t_M$ (metallicity and mass-weighted age relative to the mass trend) for trail dwarfs (red solid circles) and non-trail dwarfs (blue open circles). The dark red solid square and blue open square represent the median $\Delta$[M/H] and $\Delta t_M$ of trail and non-trail dwarfs. The trail dwarfs have higher metallicities and older ages on average than non-trail dwarfs in this plot -- a highly significant result when considered in 2D. The trail dwarfs all have similar ages and metallicities within the uncertainties, after excluding RCP~26 and DF5 as outliers. There are also mild differences between individual trail galaxies, e.g., DF2 and DF4, which suggests that their histories are more complex than being formed and quenched from identical material at identical epochs.}
\label{fig:age_metallicity}
\end{figure}

The stellar population properties for each galaxy in our sample, derived from SED fitting, are presented in Table \ref{tab:data_sedfitting}, with a summary of results in Figure~\ref{fig:mass_metallicity}. The top panel shows the stellar mass--metallicity distribution. Most of the trail dwarfs appear to follow a tight relation with a linear fit as follows (using the same methods as in Section \ref{results_cmr}):

{\footnotesize $${\rm [M/H]} = (0.12 \pm 0.16) \times (\log(M_*/M_{\odot})-7.5) - (1.25 \pm 0.11)$$}
RCP~26 and DF5 are exceptions, with outlying high and low metallicities, respectively. Non-trail dwarfs show more scatter with most of them lying below the relation for trail dwarfs. 

The mass-weighted stellar ages versus stellar masses are shown in Figure \ref{fig:mass_metallicity}, bottom panel. The trail dwarfs are consistent with all having the same ages within the uncertainties, but in more detail there is a clear systematic trend of younger ages at lower masses. We fit a mass--age relation for trail dwarfs:

{\footnotesize $$t_M = (2.92 \pm 1.46) \times (\log(M_*/M_{\odot})-7.5) - (9.06 \pm 0.81)$$}
Despite being outliers in the mass--metallicity distribution, RCP~26 and DF5 are located close to this mass--age relation. Almost all the non-trail dwarfs lie systematically below the relation.

The large uncertainties for the lower-mass (fainter) galaxies provide a warning about potential underlying systematic effects. When the S/N is low in SED fitting, the posteriors can end up mostly reflecting the priors, e.g., biasing ages towards half the age of the Universe. We have experimented with simulated data and SED fitting and found that this effect is not strong enough to remove the qualitative trends seen in Figure~\ref{fig:mass_metallicity}, but it is difficult to ``correct'' the slopes quantitatively. Instead, our focus here will be on differential comparisons between dwarfs at similar stellar masses, where any biases should be equal, and using the fitted slopes as convenient reference lines rather than having solid physical meaning.

We next define $\Delta$[M/H] or $\Delta t_M$ as the difference between the metallicity or mass-weighted age of each galaxy and the best-fit mass--metallicity or mass--age relation. In Figure \ref{fig:age_metallicity}, we present the distributions of $\Delta$[M/H] and $\Delta t_M$ for the trail and non-trail dwarfs. Although the errorbars for individual dwarfs here are large enough that most of them overlap, the trail and non-trail dwarfs as a group do exhibit systematic differences in their distributions. The median $\Delta$[M/H] shows a significant difference between the trail dwarfs (median $\Delta {\rm [M/H]}= 0.02 \pm 0.06$ dex) and non-trail dwarfs (median $\Delta {\rm [M/H]}= -0.17 \pm 0.06$ dex). The trail dwarfs have a much narrower distribution of $\Delta$[M/H] than non-trail dwarfs, as expected from the top panel of Figure \ref{fig:mass_metallicity} (if treating RCP~26 and DF5 as outliers). Also, as expected from the bottom panel of Figure \ref{fig:mass_metallicity}, trail dwarfs (median $\Delta t_M=-0.4 \pm 0.4$ Gyr) show systematically older ages than non-trail dwarfs (median $\Delta t_M=-1.4 \pm 0.3$ Gyr), again supporting different evolutionary histories for the two populations. Considering now the 2D distribution of $\Delta$[M/H] versus $\Delta t_M$, we conduct a 2D Kolmogorov--Smirnov test \citep{Fasano1987} with \texttt{ndtest}\footnote{\href{https://github.com/syrte/ndtest}{https://github.com/syrte/ndtest}} to compare the trail and non-trail galaxies. The $p$-value is 0.002, indicating that these two sub-samples cannot be drawn from the same population, thus their stellar population are significantly different. We note that the difference is dominated more by age than by metallicity. If all seven visually identified outlier galaxies in Figure~\ref{fig:age_metallicity} are removed, the 2D test still yields a high significance with a $p$-value of 0.002, although this test is less objective than using the full sample.

There could in principle be a simple explanation for these stellar population differences, since the trail dwarfs are preferentially located at smaller groupcentric radii compared to non-trail dwarfs (see Figure~\ref{fig:pa}). It is known that satellite galaxies closer to their host are on average redder and older, in connection with earlier infall times (e.g., \citealt{Venhola2019}). Examining the overall sample of dwarfs around NGC~1052, we find that both age and metallicity decline with groupcentric radius. However, after correcting for these trends, the stellar populations difference between trail and non-trail dwarfs still remains ($p=0.02$ in a 2D Kolmogorov--Smirnov test), i.e., they differ at the same projected radius.

We also show in Figure~\ref{fig:mass_metallicity} stellar population results (shaded contours) for a sample of MATLAS dwarfs using spectroscopy \citep{Heesters2023} as a reference point for the typical properties of dwarfs in massive groups. Their mass--metallicity and age--metallicity distributions appear fairly similar to that of the NGC~1052 dwarfs, with the caveats that the data and analysis methods are different, and there are biases in our results as discussed above. Note that both the MATLAS and NGC~1052 datasets are offset to lower metallicities than the Local Group dwarf relation of \citet{Kirby2013} that is commonly used for reference (an offset discussed at length by \citealt{Heesters2023}), but there are many differences in the underlying analysis methods, metallicity definitions and environments, as well as small number statistics for the Local Group dwarfs in the overlapping mass range. Also, as shown in Figure~\ref{fig:mass_metallicity}, there is no offset when comparing alternatively to the Local Group relation from \citet{Simon2019}. We emphasize that stellar populations results comparing different galaxy samples are far more reliable if homogeneous datasets and analysis methods are used (as we have done here).

In addition to the trail galaxies being older (in a mass-weighted sense) than the non-trail galaxies on average, they also quenched earlier, with median values of $\Delta t_{90} = -0.4 \pm 0.4$ and $-1.4 \pm 0.3$ Gyr ($\Delta t_{90}$ defined as the same way as $\Delta t_M$ and $\Delta$[M/H] mentioned above), respectively. On the other hand, the star formation time-scales, as measured by the parameter $t_M-t_{90}$, are similar: the median $t_M-t_{90}$ values are $0.67 \pm 0.02$ Gyr and $0.67 \pm 0.03$ Gyr for trail and non-trail dwarfs, respectively. Almost all the galaxies in our sample have a 1-$\sigma$ upper limit of $t_M-t_{90} < 2$~Gyr, which implies that the time-span from galaxy formation to quenching is less than around 4 Gyr. There is a weak hint from the modelling results that the {\it onset} of star formation is more homogeneous among the trail dwarfs than $t_M$ or $t_{90}$. This may provide an explanation for the lowest-mass trail dwarfs (DF7, RCP 16, RCP 21, RCP 28) apparently having the youngest ages ($\sim$~6--8~Gyr versus $\sim$~9--11~Gyr for the higher-mass dwarfs, assuming no mass-dependent modelling biases). It could be that the trail dwarfs all began forming stars at the same time, and then quenching by gas exhaustion or feedback was delayed at lower masses due to the low star formation efficiency.

As also noted in Section \ref{results_morph}, there are more available imaging data for the trail dwarfs than the non-trail dwarfs (especially {\it HST} observations). We find that removing the {\it HST} data from our SED fitting does not qualitatively change the conclusions above about the stellar populations.

The comparison of the stellar populations between DF2 and DF4 is important, since they are the only two galaxies along the trail known to lack DM. The crucial question is whether they originated in the same physical event or just coincidentally are two DM-free galaxies close to each other. Our SED fitting suggests that the differences in age and metallicity between DF2 and DF4 are within the 1-$\sigma$ uncertainties, which is not inconsistent with the scenario of the same origin. At the same time, DF4 is consistently redder than DF2 across multiple bands and telescopes, which does support it being older and more metal-rich. Additionally, we also find two pairs of galaxies sharing similar properties, DF9 and LEDA~4014647, and RCP~19 and RCP~20. Both galaxies in each pair have almost the same stellar mass, age, metallicity, S\'ersic index and axis ratio. Both DF9 and LEDA~4014647 have a central nucleus, and RCP~19 and RCP~20 are close to each other in the sky. These similarities might reflect a unique formation and evolution history for each pair of dwarfs, a possibility that could be explored through more detailed study. Overall, the stellar populations of the trail dwarfs are highly correlated, but not identical -- an important new ingredient in constraining formation scenarios.

\section{Assessing formation scenarios}
\label{discussion_theories}

As mentioned in Section \ref{intro}, various theories have been proposed to explain the presence of DM-deficient dwarf galaxies around the NGC~1052 galaxy group. Table \ref{tab:scores} presents these theories along with associated predictions for observable properties. Below we discuss the implications, focusing on the stellar populations (Section~\ref{sec:spops_assess}) and morphologies (Section~\ref{sec:morph_assess}), with additional properties in Section~\ref{sec:other_assess}.

\subsection{Stellar population implications}\label{sec:spops_assess}

\begin{deluxetable*}{lccccc}
\tabletypesize{\footnotesize}
\renewcommand{\arraystretch}{1.5}
\tablewidth{0pt}
\tablecaption{Comparison of different theories to explain the trail dwarfs in the NGC~1052 group. The first mark in each cell is from consideration of DF2 and DF4 alone. The second mark is from consideration of the entire trail of galaxies. The green checkmark symbol indicates that the existing observations are consistent with theoretical predictions, while the red cross symbol indicates inconsistency. The question mark signifies unclear conclusions, because the relevant observations are not available, the theoretical predictions are unclear or the evidence is mixed.
\label{tab:scores}}
\tablehead{
\colhead{} & \colhead{Bullet dwarf collision} & \colhead{Tidal stripping} & \colhead{Stellar feedback} & \colhead{Tidal dwarf} & \colhead{AGN jet or outflow}}
\startdata
    Stellar populations & \textcolor{green}{$\checkmark$} \textcolor{green}{$\checkmark$} & \textcolor{red}{$\times$} \textcolor{red}{$\times$}  & \textcolor{green}{$\checkmark$} \textcolor{red}{$\times$} & \textcolor{red}{$\times$} \textcolor{red}{$\times$} & ? ?\\
    Morphologies & \textcolor{green}{$\checkmark$} ? & \textcolor{green}{$\checkmark$} ? & \textcolor{green}{$\checkmark$} ? & \textcolor{green}{$\checkmark$} ? & ? ?\\
    Dark matter & \textcolor{green}{$\checkmark$} ? & \textcolor{green}{$\checkmark$} ? & ? ? & \textcolor{green}{$\checkmark$} ? & \textcolor{green}{$\checkmark$} ?\\
    Globular clusters & \textcolor{green}{$\checkmark$} ? & \textcolor{red}{$\times$} ? & ? ? & \textcolor{red}{$\times$} ? & \textcolor{red}{$\times$} ?\\
\enddata
\end{deluxetable*}

The basic conclusion from color--magnitude relations and stellar populations in Sections~\ref{results_cmr} and \ref{results_sed} is that the trail dwarfs seem to be a distinct population from the non-trail dwarfs, but the differences are subtle. Therefore, a formation scenario for DF2 and DF4 should explain the correlated formation of another half-dozen dwarfs, while not invoking drastically different formation and evolution physics. Most of the proposed formation scenarios have no natural explanation for the trail, which would instead have to be considered a statistical fluctuation rather than a physical association. However, even the properties of DF2 and DF4 alone could be problematic in these scenarios. In tidal formation scenarios, these galaxies were originally more massive (whether as more massive dwarfs or as debris from a giant galaxy), which should be manifested as redder and more metal-rich stars than expected for their present-day stellar masses (e.g., \citealt{Duc2000,Weilbacher2003,Zaragoza-Cardiel2024}) -- which could explain RCP~26. The observational results for DF2 and DF4 are in tension with such scenarios \citep{Buzzo2023,Gannon2023}, unless there is a conspiracy where the tidal event occurred at very high redshifts when galaxies were more metal-poor (e.g., \citealt{Naidu2022,Curti2024}), and thus ended up on the present-day red sequence after mass loss. The similarity between DF2 and DF4 also at least requires that their tidal events occur in a similar environment and at roughly the same time. As an example, \citet{Moreno2022} produced DM-deficient dwarfs from tidal stripping in a full cosmological context (FIRE-2 simulation), and these were old (4--10 Gyr) but much more metal-rich than the trail dwarfs in NGC~1052. In particular, their closest analogs to DF2 and DF4 (in size, stellar mass, velocity dispersion) are ten times more metal-rich (sitting well beyond the range shown in Figure~\ref{fig:mass_metallicity}). Also, in a tidal dwarf scenario, the galaxies would somehow have to be ejected to large distances to ensure their long-term survival with no DM (and to be consistent with the TRGB distances).

Alternatively, DF2 and DF4 could be better explained by a stellar feedback model, with stellar populations that are plausibly similar to those of other dwarfs. Expectations from an AGN jet scenario are unclear, but we cannot rule it out since the group-host NGC~1052 is identified as a low-luminosity AGN.

The bullet dwarf scenario provides the only natural explanation so far for stellar populations of the trail dwarfs to be distinct. These dwarfs are expected to have formed simultaneously from gas with a common initial metallicity, and to have quenched rapidly as their gas supply ran out. The observed dwarfs show mass dependent variations that are initially surprising in this scenario (with the caveat that some of the variation may be caused by modeling bias). However, as discussed earlier, there could be physical explanations for mass-dependent quenching times, after synchronized onsets of star formation. A mass--metallicity relation for bullet dwarf remnants might be a reflection of universal enrichment physics that is more closely linked to stellar mass than to potential well depth \citep{Baker2023}. The simulations of \citet{Lee2024} did find that more massive remnants from a bullet dwarf event may accrete more surrounding gas and sustain star formation for a longer time -- which we assume would lead to more metal enrichment. However, further theoretical work is needed with more realistic feedback models.

The GCs in the dwarfs also encode information about the overall SFHs: they likely arose in the early peak SFR, followed by the bulk of field star formation. The GCs are expected to be older and more metal-poor than the rest of the host galaxy. Small age differences for old populations are very difficult to measure, but a metallicity difference of $\sim$~0.5~dex between GCs and galaxy was estimated for DF2 and DF4 by \citet{Fensch2019a} and \citet{vanDokkum2022b}. Unfortunately, there are no clear quantitative predictions for most of the scenarios. \citet{Lee2021} did report a 0.2~dex difference from their bullet dwarf simulations, but as discussed in the next section, their feedback model may be insufficient for making detailed SFH predictions. We will discuss the GCs further in Section~\ref{sec:other_assess}.

In presenting the trail of dwarfs, \citet{vanDokkum2022a} suggested that the two galaxies at the ends, RCP~32 and DF7, could be the remnants of the progenitors in the bullet dwarf collision. In this case, their properties would differ from the rest of the trail dwarfs, as supported by their distinct morphologies. A fairly robust expectation is for them to have older and more metal-poor stars, since they would have pre-enriched and then lost the gas that formed the other dwarfs. RCP~32 is too faint for our analysis, but for DF7 we find that it follows the same mass--metallicity and mass--age trends as the rest of the trail dwarfs (Figure~\ref{fig:mass_metallicity}), implying that it is not a progenitor. The orbital configurations in the \cite{Lee2024} simulation led to one progenitor at the distant end of the trail, which would most likely correspond to RCP~32. The other progenitor's orbit has curved and no longer aligns with the trail, and hence could correspond to one of the `non-trail' dwarfs. Interestingly, DF7 was also found by \cite{Cohen2018} to be at a closer distance than DF2, based on SBF analysis, which is consistent with expectations from the trail model.

We can also revisit the timing argument from \citet{vanDokkum2022a}, where backward extrapolation of the trail galaxy positions and velocities suggested they were born together $\sim$~6--8 Gyr ago. We find that the onset of star formation for the oldest dwarfs in our sample (DF4, DF9, etc.) was $\sim$~9--11~Gyr ago. This possible mild tension should be investigated with a more rigorous study of possible trajectories, as well as with further work on the stellar populations, since several previous studies have found ages as low as 7--8 Gyr (see Appendix~\ref{stellar_pop_appendix}).

Our overall assessment of the different formation scenarios in light of the stellar population constraints is that only the bullet dwarf collision could naturally explain the distinct age and metallicity distributions for the trail of galaxies, albeit with quantitative interpretation still to be worked out. If we focus on only DF2 and DF4, then the extreme stellar feedback scenario is also possible. The AGN jet scenario may also still be considered, since there are not yet any testable predictions for stellar populations. Table~\ref{tab:scores} summarizes these conclusions in a scorecard format.

\subsection{Morphology implications}\label{sec:morph_assess}

In Section \ref{results_morph} we found that the trail dwarfs do not show significantly distinct morphological properties compared to the non-trail dwarfs, at least given our current method of dividing the two populations based only on sky positions. The trail dwarfs have their PAs surprisingly aligned with the trail, which could have a natural explanation in the bullet dwarf scenario. It is plausible that the gas clumps could be stretched along the gas trail after the collision. Subsequently, the newly formed galaxies in these gas clumps would have more stars formed along the trail and fewer in the perpendicular direction, which means that they would have prolate-spheroidal shapes with their long axes parallel to the trail. Whether or not this schematic picture works physically will require further work with simulations. Furthermore, the (weaker) presence of PA alignment among the non-trail dwarfs makes the phenomenon challenging to interpret. We also note that the PAs reported here at 1~$R_{\rm e}$ are unlikely to trace tidal effects, whose signatures have been seen in DF2 and DF4 beyond $\sim 2 R_{\rm e}$ \citep{Keim2022,Golini2024}. 

The unusually large size of DF2 was a key element in the bullet dwarf model introduced by \citet{Silk2019}, with DF4 later identified as a similar example with no DM. The low SB was explained as an effect of a high velocity collision, which leads both to low star formation efficiency and galaxy expansion. Thus there may be tension between this model and the observed heterogeneity of trail dwarf sizes, which suggests an unrelated random sample of galaxies. Further theory work is needed to see if a bullet dwarf event can produce such diverse sizes. We note some bullet dwarf simulations \citep{Lee2021} produced DM-free galaxies that were ultra-compact rather than ultra-diffuse, which might reflect feedback prescriptions that were too weak, while others \citep{Otaki2023} produced a broader range of sizes.

In summary, the PA alignments of the trail dwarfs could support the bullet dwarf scenario, although the alignments of the non-trail dwarfs would remain unexplained unless they are just a statistical fluctuation. The normalcy of the trail galaxy sizes (including scatter in SB) and shapes is difficult to understand with any exotic formation scenario where they all formed under similar conditions, most of which are expected to form LSB dwarfs. It could be that the final properties are determined by some complex interplay between the initial binding energy, the efficiency of the local feedback and tidal effects, with small variations in local conditions leading to large size variations. We conclude that there is no clear evidence from morphologies for any of the formation scenarios.

\subsection{Combining other observables}\label{sec:other_assess}

Here we review other observational constraints on formation scenarios, both existing and future, beyond those presented in this paper. Perhaps the most valuable information would be dynamical mass measurements of other trail galaxies besides DF2 and DF4, to test if they also lack DM. A less direct approach would be to conduct very deep imaging (e.g.,  \citealt{Montes2020,Keim2022,Golini2024}) to search for tidal features that would imply missing DM. Note that the tidal features around DF2 and DF4 point to the DM being missing rather than just expelled from the galaxy centers as in the stellar feedback scenario.

The unusual GC systems of DF2 and DF4 remain a key observation to explain besides their lack of DM.  The total mass fraction in GCs is unusually high -- similar to some other UDGs (e.g., \citealt{Danieli2022}) -- but the really unique properties are their unusually large sizes and high luminosities. Furthermore, the GCs have extremely homogeneous colors, suggesting a coordinated single starburst epoch involving both DF2 and DF4. The bullet dwarf scenario is so far the only one that could explain all these observations, as it involves a single, intense star formation event whose high pressures could create unusually massive GCs \citep{Silk2019,Lee2021}. A more conventional model involving galaxy mergers could also explain the unusual GCs \citep{TrujilloGomez2021} but has not been integrated self-consistently with the feedback needed to lower the DM density \citep{TrujilloGomez2022}. Neither the tidal dwarf nor the AGN outflow model is expected to produce unusual GC systems (e.g., \citealt{Fensch2019b}).  The tidal stripping scenario has been shown as capable of producing elevated numbers of GCs \citep{Ogiya2022} but not their other unusual properties. There are also preliminary indications that RCP~32 and DF9 harbor similar GCs to DF2 and DF4 \citep{Roman2021,Buzzo2023}, requiring further work from both spectroscopy and deep and high-resolution {\it HST} imaging of these and other trail dwarfs to search for and confirm unusual GCs.

For all the formation scenarios, it should be kept in mind that one or possibly more of the trail dwarfs are currently located outside of the NGC~1052 group. The absolute distances of the individual galaxies DF2 ($21.7 \pm 1.2$ Mpc) and DF4 ($20.0 \pm 1.6$ Mpc) are consistent with the distance to NGC~1052 ($20.4 \pm 1.0$~Mpc), but there is a more strongly constrained relative distance of $1.7 \pm 0.5$ Mpc between the two galaxies \citep{Danieli2020,Shen2021b,Shen2023}. Since the virial radius of the group is approximately 360 kpc \citep{Forbes2019}, this means that DF2 and DF4 cannot both be group members, and if the trail is truly a physical feature, then many of the trail members could also be outside the group. All of the formation scenarios involve dwarf galaxy formation within the group, and for them to later exit the group requires high velocities. These velocities are natural components of the bullet dwarf scenario, and can also happen with splashback after tidal stripping \citep{Moreno2022}. Furthermore, in the bullet dwarf scenario, the line-of-sight distances and velocities of the trail dwarfs are predicted to correlate linearly with their locations along the trail \citep{Lee2024}. \cite{Gannon2023} measured the velocity of DF9 as being incompatible with the trail prediction, which appears to disprove its trail membership. However, \cite{Lee2024} found individual dwarfs can deviate significantly from that relation in their simulation. All of the evidence suggests that future observations of line-of-sight distances and velocities of the trail dwarfs would be extremely valuable (e.g., M.A.~Keim et al., in preparation).

Considering the preceding point, and all of the available observational constraints summarized in Table~\ref{tab:scores}, we can now provide an overall assessment of the formation scenarios, while keeping in mind that all of them except the extreme feedback scenario are certain to occur in reality (even if rarely). The properties of DF2 and DF4 alone are best explained by the bullet dwarf collision.  Stellar feedback does fairly well in explaining these galaxies, with some areas of potential tension relating to their DM content and GC systems.  The other scenarios all have at least one type of observation that is very difficult to explain. Folding in the correlated properties of all the trail galaxies strengthens further the support for the bullet dwarf scenario, as none of the other scenarios naturally produces such a trail (except one might speculate that collimated AGN feedback could produce a linear string of dwarfs).

Despite the success so far of the bullet dwarf scenario, there are still questions to answer about some of the trends in stellar populations and morphologies, and it is worth continuing to consider alternative explanations. One relevant observation is the widespread finding in galaxy groups, including MATLAS, of satellite subsystems in planar or linear configurations (e.g., \citealt{Heesters2021}). These features are often aligned with large-scale structure beyond the group, which \citet{vanDokkum2022a} determined was {\it not} the case for the NGC~1052 trail. Also, it is expected that alignments with large-scale filaments would produce the {\it opposite} effect to the trail dwarf observations, with the photometric minor axis corresponding to the angular momentum axis that is expected to be aligned with a filament \citep{Rong2020}. Furthermore, it is not clear why there would be a connection between planes of satellites and DM-free dwarfs. These complexities could be explored in the future with in-depth analysis of predictions from cosmological simulations (e.g., \citealt{Muller2024}), while incorporating the boundary conditions of two DM-deficient UDGs with unusual GC systems. A specific scenario can also be entertained wherein NGC~1035 hosts a second galaxy group that is either merging with the NGC~1052 or overlaps in projection (see Figure~\ref{fig:pa}), which has been discussed in previous studies (e.g., \citealt{Trujillo2019,Roman2021,Shen2021b}).  This juxtaposition might lead to the illusion of a trail with distinctive dwarfs. Additional measurements of galaxy distances and velocities could provide strong tests of this picture.

We note that although the bullet-dwarf collision scenario appears to be the best explanation for DF2, DF4, and their associated trail dwarfs, this is so far a unique system. Other types of DM-deficient dwarfs may well exist that are produced by different mechanisms as discussed above. Furthermore, there is a population of isolated gas-rich UDGs found to have unexpectedly small (but non-zero) amounts of DM (e.g., \citealt{Kong2022,ManceraPina2024}), and the origins of these galaxies are entirely unclear.

\section{Conclusions}
\label{conclusions}

In this work, we test the bullet dwarf collision theory proposed in \cite{vanDokkum2022a}, along with other scenarios in the literature, by studying the stellar populations and morphologies of the NGC~1052 trail dwarf galaxies and other galaxies in the vicinity (non-trail dwarfs), based on new observations from the {\it Hubble Space Telescope} combined with existing imaging from the $u$ band to mid-IR.

We do not find that the trail dwarfs have larger sizes overall than non-trail dwarfs as originally envisioned in the bullet dwarf scenario developed around DF2 and DF4. The photometric position angles of most trail dwarfs show parallel alignment to the trail itself, suggesting that the trail dwarfs might be physically associated. However, the reason why many non-trail dwarfs have their PAs aligned with the trail structure as well is unknown. Other morphological parameters do not show significant difference between the two groups of dwarfs.

Based on our SED fitting with {\tt PROSPECTOR}, we find that the trail dwarfs have significantly different stellar population properties compared to the non-trail dwarfs, with older ages and higher metallicities on average.

A few of the dwarfs that are apparently part of the trail may actually be interlopers, which could be recognized by their properties deviating from the rest of the trail dwarfs. RCP~26 is the strongest outlier, with its high metallicity, flattened shape and misaligned position angle. DF7 and RCP~16 are more flattened than other trail dwarfs, and DF5 is a low-metallicity outlier. More information including distance and velocity is needed to confirm their membership in the trail.

In summary, based on the observational results and theoretical work so far, we see that the bullet dwarf collision scenario exhibits the most potential for simultaneously explaining DF2, DF4, and other dwarfs in the sky region near NGC~1052, compared to other proposed scenarios. Future follow-up measurements of distance, velocity, and DM content are crucial, providing more direct tests of the bullet dwarf collision scenario and of the existence of the trail structure.

\begin{acknowledgments}

We would like to express our deepest gratitude to Tom Jarrett, who passed away after the submission of this manuscript, for his invaluable contributions and dedication to this research. We thank Duncan Forbes, Joel Leja and Eun-jin Shin for helpful discussions, and Yotam Cohen and the rest of the STScI ACS team for their help with the data reduction. We are grateful to the anonymous referee for numerous suggestions that clarified our results. Supported by National Science Foundation grant AST-2308390. Support for Program number HST-GO-16912 was provided through a grant from the STScI under NASA contract NAS5-26555. This work was supported by a NASA Keck PI Data Award, administered by the NASA Exoplanet Science Institute. Data presented herein were obtained at the W.~M.\ Keck Observatory from telescope time allocated to the National Aeronautics and Space Administration through the agency’s scientific partnership with the California Institute of Technology and the University of California. The Observatory was made possible by the generous financial support of the W.~M.\ Keck Foundation. The authors wish to recognize and acknowledge the very significant cultural role and reverence that the summit of Maunakea has always had within the indigenous Hawaiian community. We are most fortunate to have the opportunity to conduct observations from this mountain. This work was authored by an employee of Caltech/IPAC under Contract No. 80GSFC21R0032 with the National Aeronautics and Space Administration. This work is based in part on observations made with the {\it Spitzer Space Telescope}, which was operated by the Jet Propulsion Laboratory, California Institute of Technology under a contract with NASA.

\end{acknowledgments}

\bibliography{ms}{}
\bibliographystyle{aasjournal}

\restartappendixnumbering
\appendix

\section{Photometric modeling details and tests}
\label{morph_appendix}

\begin{figure}
\centering
\includegraphics[width=0.47\textwidth]{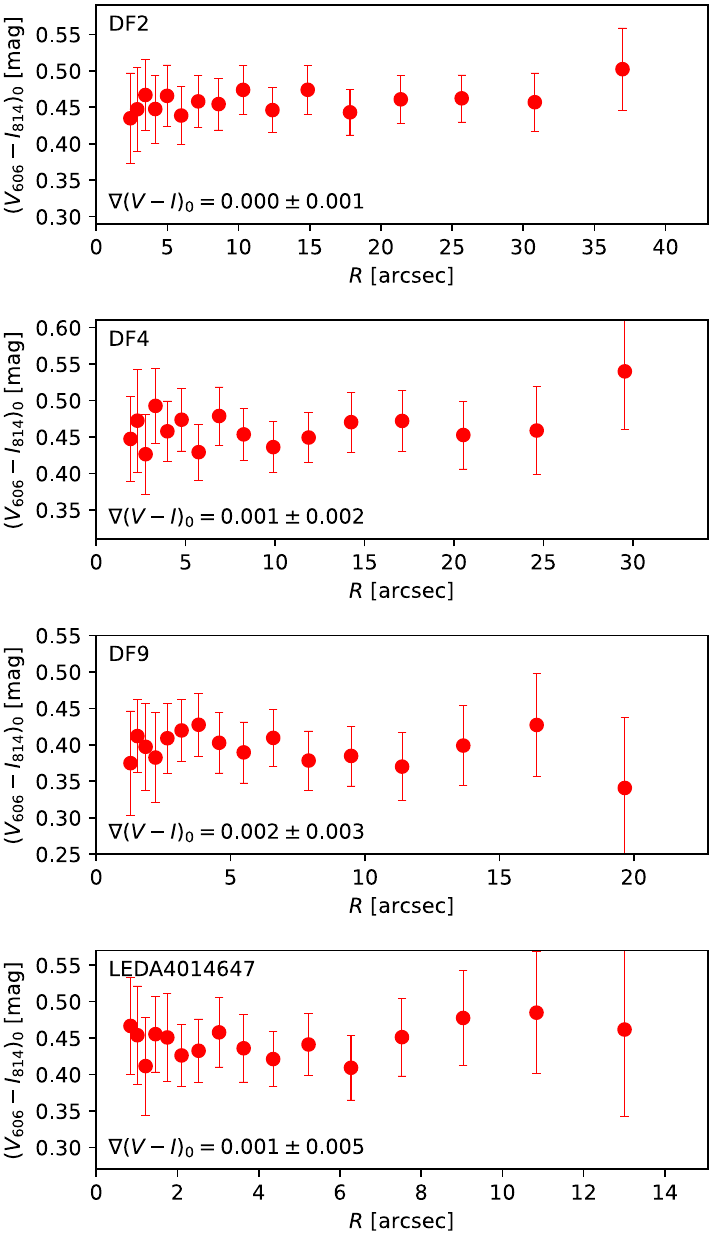}
\caption{{\it HST} $V_{606}-I_{814}$ color profiles within 2$R_{\rm e}$ of the four brightest trail dwarfs, DF2, DF4, DF9 and LEDA~4014647. The x-axis shows the semi-major axis radii. The calculated color gradients are listed in the lower left corner of each panel, and all these values are consistent with zero.}
\label{fig:color_gradient}
\end{figure}

\begin{figure*}
\centering
\includegraphics[width=1\textwidth]{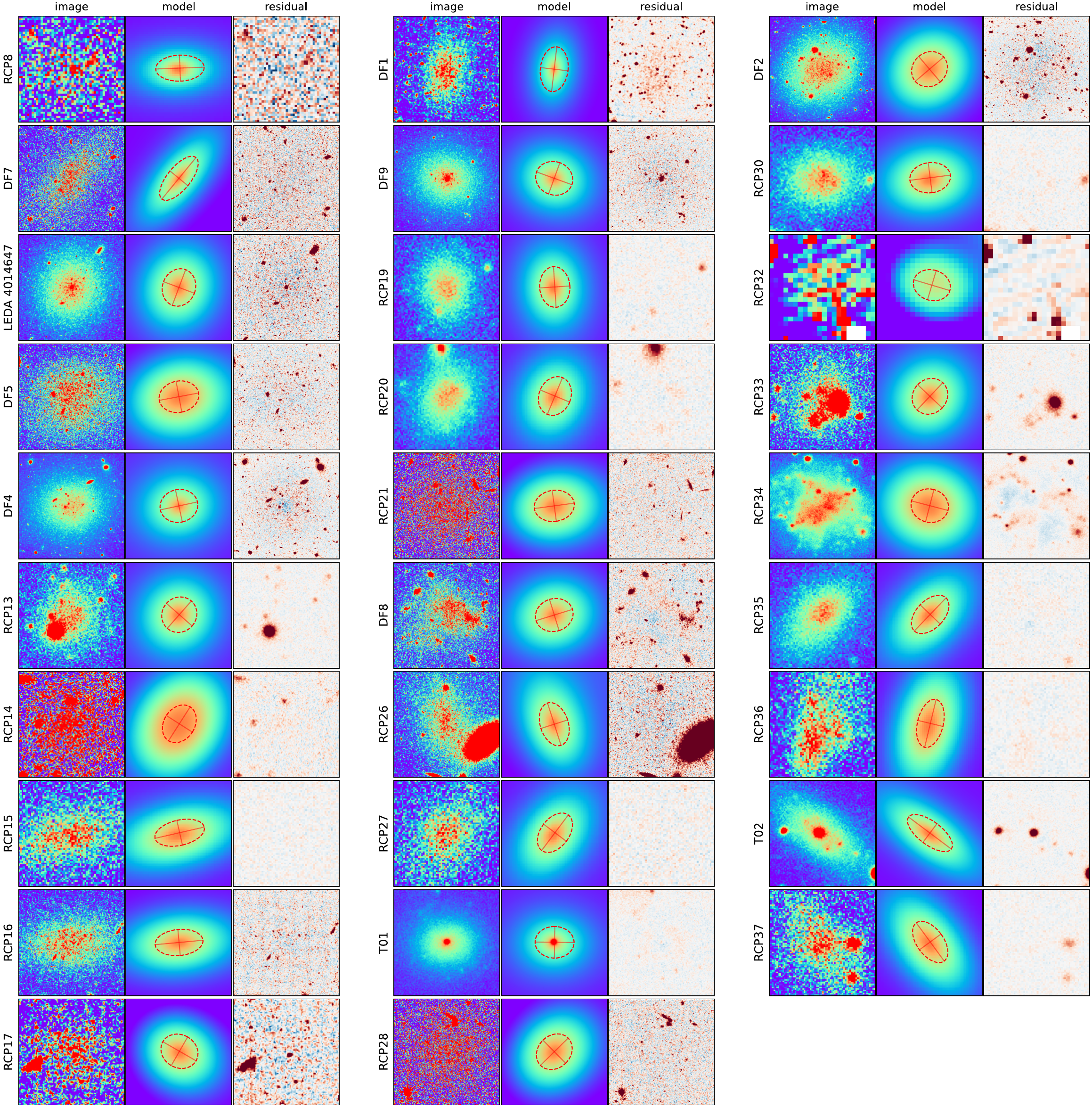}
\vspace{-10pt}
\caption{GALFIT fitting of the stacked image of two {\it HST} bands for all the galaxies with {\it HST} observations, and the stacked image of DECaLS with the $g$ and $r$ bands for the other galaxies. The fitting of RCP~17 and RCP~32 is based on their rebinned {\it HST} stacked images. The galaxies are listed from top to bottom and from left to right according to right ascension. For the three panels of each galaxy, we show the original image, the best-fitting model, and the residual, respectively. The cutout images here have sizes of 3 times the circularized effective radius of each galaxy, while the actual input images to GALFIT typically have a side length of about 8 times the effective radius. The red dashed ellipses represent the isophote at $R_{\rm e}$, and the red solid lines show corresponding major and minor axes.}
\label{fig:galfit_all}
\end{figure*}

\begin{figure*}
\centering
\includegraphics[width=1\textwidth]{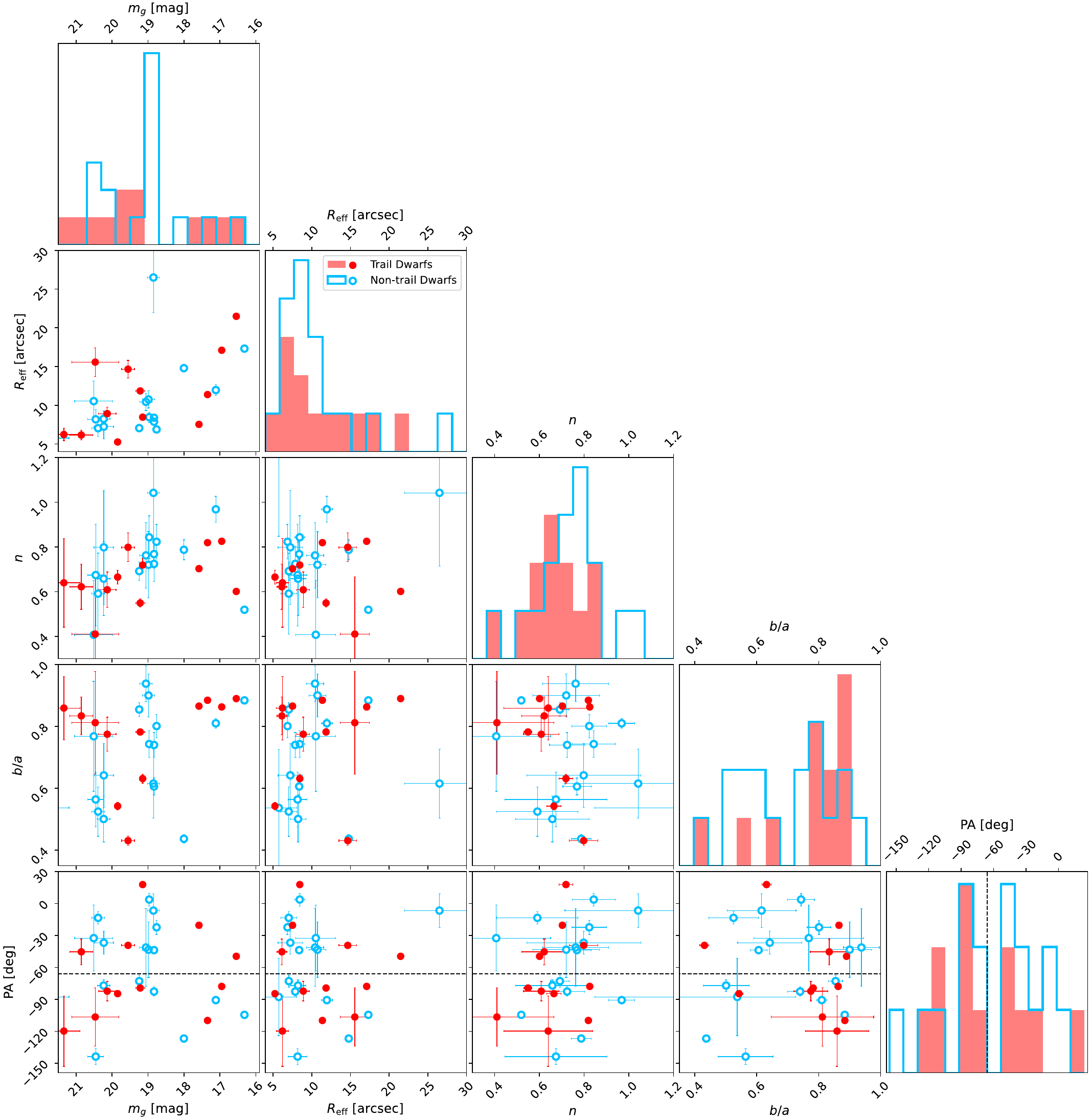}
\vspace{-10pt}
\caption{One-dimensional (histograms) and two-dimensional distributions for the DECaLS $g$ band magnitude and four morphological parameters from GALFIT of the trail dwarfs (red) and non-trail dwarfs (blue). Note that the PA values wrap around, with 60 and $-120$ degrees being equivalent. In the bottom panel, the black dashed lines represent the trail PA.}
\label{fig:morphology_all}
\end{figure*}

We show the {\it HST} $V_{606}-I_{814}$ color profiles of the four brightest trail dwarfs in Figure \ref{fig:color_gradient}. Within 2$R_{\rm e}$, all four dwarfs have color gradients consistent with zero within the uncertainties.

Figure \ref{fig:galfit_all} shows the GALFIT fitting results with the image with the highest signal-to-noise ratio for each galaxy. For all the galaxies with {\it HST} observations, this means the stacked image of the two {\it HST} bands. For the other galaxies, we use the stacked DECaLS image with the $g$ and $r$ bands. In Figure \ref{fig:morphology_all}, we present a corner plot of the morphological parameters from GALFIT of all the galaxies in our sample.

\begin{figure*}
\centering
\includegraphics[width=0.48\textwidth]{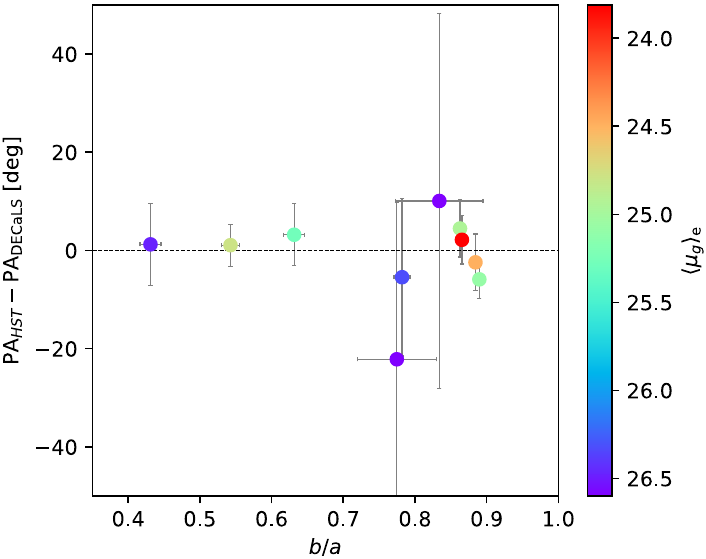}
\caption{The PA measurement difference between {\it HST} and DECaLS of trail dwarfs, and how it relates to the axis ratio ($x$-axis) and mean SB in the $g$ band (see bar on the right for color-coding).}
\label{fig:pa_comparison}
\end{figure*}

In Section \ref{results_morph}, we discuss the distribution of PAs for the dwarf galaxies. Some of these measurements are from {\it HST} images, while others are from DECaLS images. Here we compare the measurements from different images for consistency. Figure \ref{fig:pa_comparison} shows the differences in PA measurements for trail dwarfs between {\it HST} and DECaLS, and how these differences vary with axis ratio and mean SB. We find that accurate PA measurements are challenging when the galaxy has low SB ($\langle\mu_g\rangle_\mathrm{e} > 26$) and is very round ($b/a > 0.8$). For all of the trail dwarfs, the PAs from {\it HST} and DECaLS are consistent within the uncertainties. For the non-trail dwarfs, which generally have $b/a < 0.8$, we expect that their PA measurements from DECaLS will also be reliable.

\cite{Roman2021} investigated the morphologies of LSB galaxies in the NGC~1052 group with DECaLS Data Release 7 imaging, which is shallower than the data we use. For all the galaxies in our sample that overlap with \cite{Roman2021}, our measurements agree well with their results. \cite{Keim2022} made use of Dragonfly deep imaging of DF2 and DF4 and found strong twists of their PAs with radius. Our PA measurements for these two galaxies are consistent with their results at $1 R_{\rm e}$.

The final photometric measurements (magnitudes and colors, with uncertainties) of our sample galaxies are reported in Table~\ref{tab:data_photometry}.

\begin{deluxetable*}{lccccccccccccccc}
\tabletypesize{\scriptsize}
\renewcommand{\arraystretch}{1.1}
\tablewidth{0pt}
\tablecaption{Photometry of the dwarf galaxies in our sample, sorted by right ascension.
\label{tab:data_photometry}}
\tablehead{
\colhead{} & \multicolumn{3}{c}{{\it HST} ACS/WFC} & \multicolumn{3}{c}{CFHT MegaCam} & \multicolumn{4}{c}{DECaLS} & \multicolumn{1}{c}{{\it Spitzer}} & \multicolumn{4}{c}{{\it WISE}}\\
\cmidrule(rl){2-4} \cmidrule(rl){5-7} \cmidrule(rl){8-11} \cmidrule(rl){12-12} \cmidrule(rl){13-16}
Galaxy & $V_{606}$ & $I_{814}$ & $V_{606}-I_{814}$ & $g$ & $u-g$ & $g-i$ & $g$ & $g-r$ & $g-i$ & $g-z$ & $3.6\mu$m & W1 & W2 & W3 & W4}
\startdata
    RCP 8 & -- & -- & -- & -- & -- & -- & 21.79 & 0.58 & 0.79 & 0.98 & -- & -- & -- & -- & -- \\
    \specialrule{0em}{-6pt}{1pt}
    {} & {} & {} & {} & {} & {} & {} & {\tiny $\pm$ 0.60} & {\tiny $\pm$ 0.65} & {\tiny $\pm$ 0.78} & {\tiny $\pm$ 0.70} & {} & {} & {}\\
    DF7 (RCP 9) & 19.18 & 18.75 & 0.43 & -- & -- & -- & 19.46 & 0.51 & 0.72 & 0.91 & -- & 19.34 & 19.63 & $>$18.09 & $>$16.77 \\
    \specialrule{0em}{-6pt}{1pt}
    {} & {\tiny $\pm$ 0.08} & {\tiny $\pm$ 0.08} & {\tiny $\pm$ 0.04} & {} & {} & {} & {\tiny $\pm$ 0.18} & {\tiny $\pm$ 0.20} & {\tiny $\pm$ 0.22} & {\tiny $\pm$ 0.29} & {} & {\tiny $\pm$ 0.92} & {\tiny $\pm$ 0.86}\\
    LEDA 4014647 & 17.14 & 16.72 & 0.42 & 17.66 & 1.03 & 0.92 & 17.58 & 0.60 & 0.82 & 0.92 & -- & 17.66 & 18.32 & $>$18.94 & $>$17.32 \\
    \specialrule{0em}{-6pt}{1pt}
    {} & {\tiny $\pm$ 0.01} & {\tiny $\pm$ 0.01} & {\tiny $\pm$ 0.01} & {\tiny $\pm$ 0.02} & {\tiny $\pm$ 0.04} & {\tiny $\pm$ 0.02} & {\tiny $\pm$ 0.02} & {\tiny $\pm$ 0.03} & {\tiny $\pm$ 0.03} & {\tiny $\pm$ 0.04} & {} & {\tiny $\pm$ 0.08} & {\tiny $\pm$ 0.45}\\
    DF4 (RCP 11) & 16.50 & 16.04 & 0.46 & 17.03 & 1.26 & 0.94 & 16.95 & 0.64 & 0.89 & 1.01 & 16.86 & 16.78 & 17.52 & $>$17.44 & $>$16.18 \\
    \specialrule{0em}{-6pt}{1pt}
    {} & {\tiny $\pm$ 0.01} & {\tiny $\pm$ 0.01} & {\tiny $\pm$ 0.01} & {\tiny $\pm$ 0.02} & {\tiny $\pm$ 0.05} & {\tiny $\pm$ 0.04} & {\tiny $\pm$ 0.02} & {\tiny $\pm$ 0.03} & {\tiny $\pm$ 0.03} & {\tiny $\pm$ 0.04} & {\tiny $\pm$ 0.02} & {\tiny $\pm$ 0.07} & {\tiny $\pm$ 0.41}\\
    DF5 (RCP 12) & 18.82 & 18.47 & 0.35 & 19.23 & 1.06 & 0.81 & 19.22 & 0.51 & 0.76 & 0.86 & 19.65 & 19.51 & 20.06 & $>$18.36 & $>$17.05 \\
    \specialrule{0em}{-6pt}{1pt}
    {} & {\tiny $\pm$ 0.02} & {\tiny $\pm$ 0.02} & {\tiny $\pm$ 0.02} & {\tiny $\pm$ 0.11} & {\tiny $\pm$ 0.19} & {\tiny $\pm$ 0.14} & {\tiny $\pm$ 0.13} & {\tiny $\pm$ 0.17} & {\tiny $\pm$ 0.19} & {\tiny $\pm$ 0.27} & {\tiny $\pm$ 0.11} & {\tiny $\pm$ 0.51} & {\tiny $\pm$ 0.88}\\
    RCP 13 & -- & -- & -- & -- & -- & -- & 19.06 & 0.54 & 0.78 & 0.88 & -- & 19.03 & $>$19.36 & $>$18.04 & $>$16.71 \\
    \specialrule{0em}{-6pt}{1pt}
    {} & {} & {} & {} & {} & {} & {} & {\tiny $\pm$ 0.17} & {\tiny $\pm$ 0.12} & {\tiny $\pm$ 0.13} & {\tiny $\pm$ 0.19} & {} & {\tiny $\pm$ 0.56} \\
    RCP 14 & -- & -- & -- & -- & -- & -- & 20.51 & 0.27 & 0.55 & 0.48 & -- & -- & -- & -- & -- \\
    \specialrule{0em}{-6pt}{1pt}
    {} & {} & {} & {} & {} & {} & {} & {\tiny $\pm$ 0.53} & {\tiny $\pm$ 0.51} & {\tiny $\pm$ 0.55} & {\tiny $\pm$ 0.84} & {} & {} & {}\\
    RCP 15 & -- & -- & -- & -- & -- & -- & 20.24 & 0.47 & 0.59 & 0.63 & -- & 21.21 & $>$19.36 & $>$18.58 & $>$17.08 \\
    \specialrule{0em}{-6pt}{1pt}
    {} & {} & {} & {} & {} & {} & {} & {\tiny $\pm$ 0.18} & {\tiny $\pm$ 0.18} & {\tiny $\pm$ 0.23} & {\tiny $\pm$ 0.31} & {} & {\tiny $\pm$ 0.87} \\
    RCP 16 (Ta21-12000) & 19.41 & 19.03 & 0.38 & 19.89 & 1.12 & 0.85 & 19.85 & 0.61 & 0.80 & 0.90 & -- & 20.06 & 20.46 & $>$19.28 & $>$17.90 \\
    \specialrule{0em}{-6pt}{1pt}
    {} & {\tiny $\pm$ 0.04} & {\tiny $\pm$ 0.04} & {\tiny $\pm$ 0.03} & {\tiny $\pm$ 0.11} & {\tiny $\pm$ 0.18} & {\tiny $\pm$ 0.13} & {\tiny $\pm$ 0.09} & {\tiny $\pm$ 0.10} & {\tiny $\pm$ 0.11} & {\tiny $\pm$ 0.15} & {} & {\tiny $\pm$ 0.44} & {\tiny $\pm$ 0.86}\\
    RCP 17 & 20.84 & 20.58 & 0.26 & 21.59 & 0.99 & 0.81 & 21.34 & 0.35 & 0.63 & 0.70 & -- & -- & -- & -- & -- \\
    \specialrule{0em}{-6pt}{1pt}
    {} & {\tiny $\pm$ 0.18} & {\tiny $\pm$ 0.20} & {\tiny $\pm$ 0.15} & {\tiny $\pm$ 0.58} & {\tiny $\pm$ 0.95} & {\tiny $\pm$ 0.76} & {\tiny $\pm$ 0.45} & {\tiny $\pm$ 0.65} & {\tiny $\pm$ 0.64} & {\tiny $\pm$ 0.95} \\
    DF1 (RCP 18) & 18.28 & 17.80 & 0.48 & 18.84 & 0.84 & 0.82 & 18.85 & 0.62 & 0.87 & 1.15 & -- & 18.49 & 18.46 & $>$17.10 & $>$15.92 \\
    \specialrule{0em}{-6pt}{1pt}
    {} & {\tiny $\pm$ 0.09} & {\tiny $\pm$ 0.09} & {\tiny $\pm$ 0.08} & {\tiny $\pm$ 0.12} & {\tiny $\pm$ 0.19} & {\tiny $\pm$ 0.17} & {\tiny $\pm$ 0.16} & {\tiny $\pm$ 0.20} & {\tiny $\pm$ 0.22} & {\tiny $\pm$ 0.30} & {} & {\tiny $\pm$ 0.37} & {\tiny $\pm$ 0.82}\\
    DF9 & 16.87 & 16.47 & 0.40 & 17.40 & 1.12 & 0.90 & 17.34 & 0.61 & 0.84 & 0.94 & -- & 17.35 & 17.99 & $>$17.76 & $>$16.45\\
    \specialrule{0em}{-6pt}{1pt}
    {} & {\tiny $\pm$ 0.01} & {\tiny $\pm$ 0.01} & {\tiny $\pm$ 0.01} & {\tiny $\pm$ 0.03} & {\tiny $\pm$ 0.05} & {\tiny $\pm$ 0.04} & {\tiny $\pm$ 0.03} & {\tiny $\pm$ 0.03} & {\tiny $\pm$ 0.03} & {\tiny $\pm$ 0.04} & {} & {\tiny $\pm$ 0.07} & {\tiny $\pm$ 0.36}\\
    RCP 19 & -- & -- & -- & 18.97 & 1.09 & 0.75 & 18.97 & 0.52 & 0.75 & 0.80 & -- & 19.16 & 19.89 & $>$18.75 & $>$17.03 \\
    \specialrule{0em}{-6pt}{1pt}
    {} & {} & {} & {} & {\tiny $\pm$ 0.11} & {\tiny $\pm$ 0.13} & {\tiny $\pm$ 0.08} & {\tiny $\pm$ 0.10} & {\tiny $\pm$ 0.07} & {\tiny $\pm$ 0.08} & {\tiny $\pm$ 0.11} & {} & {\tiny $\pm$ 0.29} & {\tiny $\pm$ 0.89} \\
    RCP 20 & -- & -- & -- & -- & -- & -- & 18.76 & 0.56 & 0.74 & 0.88 & -- & 18.95 & 19.59 & $>$18.85 & $>$16.95 \\
    \specialrule{0em}{-6pt}{1pt}
    {} & {} & {} & {} & {} & {} & {} & {\tiny $\pm$ 0.07} & {\tiny $\pm$ 0.06} & {\tiny $\pm$ 0.07} & {\tiny $\pm$ 0.09} & {} & {\tiny $\pm$ 0.20} & {\tiny $\pm$ 0.70} \\
    RCP 21 & 19.73 & 19.35 & 0.37 & 20.13 & 0.97 & 0.89 & 20.14 & 0.54 & 0.71 & 0.96 & -- & 19.83 & 20.54 & $>$18.79 & $>$16.98 \\
    \specialrule{0em}{-6pt}{1pt}
    {} & {\tiny $\pm$ 0.13} & {\tiny $\pm$ 0.14} & {\tiny $\pm$ 0.07} & {\tiny $\pm$ 0.24} & {\tiny $\pm$ 0.37} & {\tiny $\pm$ 0.26} & {\tiny $\pm$ 0.24} & {\tiny $\pm$ 0.24} & {\tiny $\pm$ 0.31} & {\tiny $\pm$ 0.38} & {} & {\tiny $\pm$ 0.50} & {\tiny $\pm$ 1.03}\\
    DF8 (RCP 24) & 19.06 & 18.85 & 0.21 & -- & -- & -- & 19.24 & 0.29 & 0.40 & 0.40 & -- & 20.06 & 21.11 & $>$18.69 & $>$17.12 \\
    \specialrule{0em}{-6pt}{1pt}
    {} & {\tiny $\pm$ 0.04} & {\tiny $\pm$ 0.04} & {\tiny $\pm$ 0.03} & {} & {} & {} & {\tiny $\pm$ 0.06} & {\tiny $\pm$ 0.08} & {\tiny $\pm$ 0.10} & {\tiny $\pm$ 0.15} & {} & {\tiny $\pm$ 0.61} & {\tiny $\pm$ 1.03} \\
    RCP 26 & 18.58 & 18.12 & 0.47 & 19.20 & 1.30 & 1.01 & 19.15 & 0.70 & 0.98 & 1.11 & -- & 18.84 & 19.64 & $>$18.44 & $>$17.88 \\
    \specialrule{0em}{-6pt}{1pt}
    {} & {\tiny $\pm$ 0.04} & {\tiny $\pm$ 0.04} & {\tiny $\pm$ 0.02} & {\tiny $\pm$ 0.08} & {\tiny $\pm$ 0.14} & {\tiny $\pm$ 0.09} & {\tiny $\pm$ 0.08} & {\tiny $\pm$ 0.09} & {\tiny $\pm$ 0.09} & {\tiny $\pm$ 0.11} & {} & {\tiny $\pm$ 0.20} & {\tiny $\pm$ 0.75}\\
    RCP 27 & -- & -- & -- & -- & -- & -- & 20.23 & 0.52 & 0.73 & 0.65 & -- & 20.79 & 21.15 & $>$18.87 & $>$17.30 \\
    \specialrule{0em}{-6pt}{1pt}
    {} & {} & {} & {} & {} & {} & {} & {\tiny $\pm$ 0.27} & {\tiny $\pm$ 0.18} & {\tiny $\pm$ 0.22} & {\tiny $\pm$ 0.34} & {} & {\tiny $\pm$ 0.86} & {\tiny $\pm$ 0.99} \\
    T01 & -- & -- & -- & 17.18 & 1.27 & 0.91 & 17.11 & 0.63 & 0.88 & 1.02 & -- & 16.74 & 17.44 & $>$17.17 & $>$16.26 \\
    \specialrule{0em}{-6pt}{1pt}
    {} & {} & {} & {} & {\tiny $\pm$ 0.14} & {\tiny $\pm$ 0.28} & {\tiny $\pm$ 0.18} & {\tiny $\pm$ 0.06} & {\tiny $\pm$ 0.02} & {\tiny $\pm$ 0.02} & {\tiny $\pm$ 0.03} & {} & {\tiny $\pm$ 0.07} & {\tiny $\pm$ 0.13} \\
    RCP 28 & 20.42 & 20.05 & 0.37 & 20.88 & 0.99 & 0.84 & 20.86 & 0.60 & 0.84 & 0.82 & -- & 20.84 & 21.97 & $>$19.18 & $>$17.04 \\
    \specialrule{0em}{-6pt}{1pt}
    {} & {\tiny $\pm$ 0.13} & {\tiny $\pm$ 0.13} & {\tiny $\pm$ 0.08} & {\tiny $\pm$ 0.36} & {\tiny $\pm$ 0.47} & {\tiny $\pm$ 0.39} & {\tiny $\pm$ 0.32} & {\tiny $\pm$ 0.39} & {\tiny $\pm$ 0.42} & {\tiny $\pm$ 0.56} & {} & {\tiny $\pm$ 0.75} & {\tiny $\pm$ 1.04}\\
    DF2 (RCP 29) & 16.11 & 15.66 & 0.44 & 16.60 & 1.18 & 0.90 & 16.55 & 0.61 & 0.84 & 0.98 & 16.46 & 16.40 & 17.06 & $>$17.51 & $>$15.94 \\
    \specialrule{0em}{-6pt}{1pt}
    {} & {\tiny $\pm$ 0.01} & {\tiny $\pm$ 0.01} & {\tiny $\pm$ 0.01} & {\tiny $\pm$ 0.02} & {\tiny $\pm$ 0.04} & {\tiny $\pm$ 0.03} & {\tiny $\pm$ 0.02} & {\tiny $\pm$ 0.03} & {\tiny $\pm$ 0.03} & {\tiny $\pm$ 0.03} & {\tiny $\pm$ 0.01} & {\tiny $\pm$ 0.05} & {\tiny $\pm$ 0.24}\\
    RCP 30 & -- & -- & -- & -- & -- & -- & 18.83 & 0.55 & 0.75 & 0.84 & -- & 19.02 & 19.39 & $>$18.81 & $>$17.31 \\
    \specialrule{0em}{-6pt}{1pt}
    {} & {} & {} & {} & {} & {} & {} & {\tiny $\pm$ 0.08} & {\tiny $\pm$ 0.07} & {\tiny $\pm$ 0.08} & {\tiny $\pm$ 0.10} & {} & {\tiny $\pm$ 0.24} & {\tiny $\pm$ 0.73} \\
    RCP 32 & 19.97 & 19.47 & 0.50 & 20.19 & 0.91 & 0.78 & 20.47 & 0.54 & 0.79 & 0.43 & -- & -- & -- & -- & -- \\
    \specialrule{0em}{-6pt}{1pt}
    {} & {\tiny $\pm$ 0.21} & {\tiny $\pm$ 0.22} & {\tiny $\pm$ 0.16} & {\tiny $\pm$ 0.48} & {\tiny $\pm$ 0.59} & {\tiny $\pm$ 0.45} & {\tiny $\pm$ 0.66} & {\tiny $\pm$ 0.48} & {\tiny $\pm$ 0.73} & {\tiny $\pm$ 1.01} \\
    RCP 33 & -- & -- & -- & -- & -- & -- & 18.99 & 0.50 & 0.73 & 0.88 & -- & 19.29 & 19.84 & $>$17.04 & $>$16.47 \\
    \specialrule{0em}{-6pt}{1pt}
    {} & {} & {} & {} & {} & {} & {} & {\tiny $\pm$ 0.17} & {\tiny $\pm$ 0.14} & {\tiny $\pm$ 0.17} & {\tiny $\pm$ 0.22} & {} & {\tiny $\pm$ 0.39} & {\tiny $\pm$ 0.84} \\
    RCP 34 & -- & -- & -- & -- & -- & -- & 16.32 & 0.32 & 0.43 & 0.48 & -- & 16.76 & 17.08 & $>$16.62 & $>$16.09 \\
    \specialrule{0em}{-6pt}{1pt}
    {} & {} & {} & {} & {} & {} & {} & {\tiny $\pm$ 0.02} & {\tiny $\pm$ 0.02} & {\tiny $\pm$ 0.02} & {\tiny $\pm$ 0.03} & {} & {\tiny $\pm$ 0.15} & {\tiny $\pm$ 0.45} \\
    RCP 35 & -- & -- & -- & 18.86 & 1.02 & 0.77 & 18.83 & 0.53 & 0.71 & 0.81 & -- & 18.99 & 19.90 & $>$18.73 & $>$16.91 \\
    \specialrule{0em}{-6pt}{1pt}
    {} & {} & {} & {} & {\tiny $\pm$ 0.07} & {\tiny $\pm$ 0.11} & {\tiny $\pm$ 0.06} & {\tiny $\pm$ 0.07} & {\tiny $\pm$ 0.05} & {\tiny $\pm$ 0.06} & {\tiny $\pm$ 0.09} & {} & {\tiny $\pm$ 0.28} & {\tiny $\pm$ 0.77} \\
    RCP 36 & -- & -- & -- & 20.35 & 0.92 & 0.51 & 20.39 & 0.47 & 0.56 & 0.68 & -- & 20.55 & 21.19 & $>$19.22 & $>$17.66 \\
    \specialrule{0em}{-6pt}{1pt}
    {} & {} & {} & {} & {\tiny $\pm$ 0.21} & {\tiny $\pm$ 0.23} & {\tiny $\pm$ 0.19} & {\tiny $\pm$ 0.18} & {\tiny $\pm$ 0.15} & {\tiny $\pm$ 0.20} & {\tiny $\pm$ 0.34} & {} & {\tiny $\pm$ 0.69} & {\tiny $\pm$ 0.99} \\
    T02 & -- & -- & -- & -- & -- & -- & 18.00 & 0.60 & 0.81 & 0.91 & -- & 17.87 & 18.45 & $>$18.46 & $>$17.09 \\
    \specialrule{0em}{-6pt}{1pt}
    {} & {} & {} & {} & {} & {} & {} & {\tiny $\pm$ 0.05} & {\tiny $\pm$ 0.04} & {\tiny $\pm$ 0.05} & {\tiny $\pm$ 0.07} & {} & {\tiny $\pm$ 0.13} & {\tiny $\pm$ 0.49} \\
    RCP 37 & -- & -- & -- & -- & -- & -- & 20.46 & 0.64 & 0.84 & 0.96 & -- & 20.78 & 21.43 & $>$19.29 & $>$17.35 \\
    \specialrule{0em}{-6pt}{1pt}
    {} & {} & {} & {} & {} & {} & {} & {\tiny $\pm$ 0.22} & {\tiny $\pm$ 0.17} & {\tiny $\pm$ 0.21} & {\tiny $\pm$ 0.34} & {} & {\tiny $\pm$ 0.89} & {\tiny $\pm$ 0.91} \\
\enddata
\tablecomments{
(1) The magnitudes shown in the table are apparent AB magnitudes after correcting for Galactic extinction. (2) `–' represents unavailable data. (3) `$>$' stands for the 1-$\sigma$ lower limit magnitudes.}
\end{deluxetable*}

\section{Stellar Populations Tests}
\label{stellar_pop_appendix}

\begin{figure*}
\centering
\includegraphics[width=0.8\textwidth]{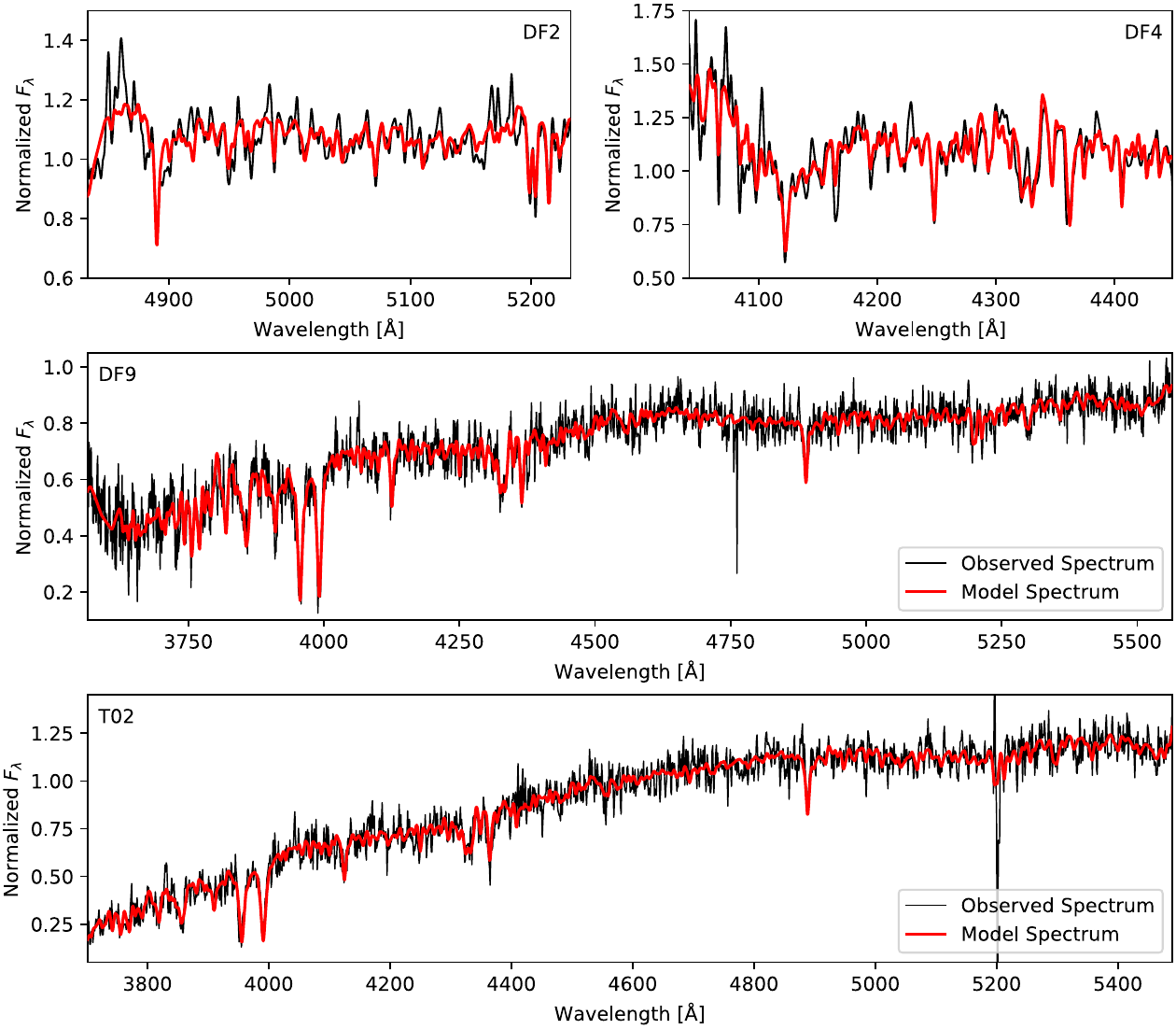}
\caption{KCWI spectra of DF2, DF4, DF9 and T02 (black lines) and the best fitting models from {\tt PROSPECTOR} (red lines; joint spectral and SED fits). The DF2 and DF4 spectra have been smoothed to a spectral resolution of 2.5 \AA\ to match the resolution of the MILES templates used in the fitting.}
\label{fig:df2_df4_df9_spec}
\end{figure*}

\begin{figure*}
\centering
\includegraphics[width=0.7\textwidth]{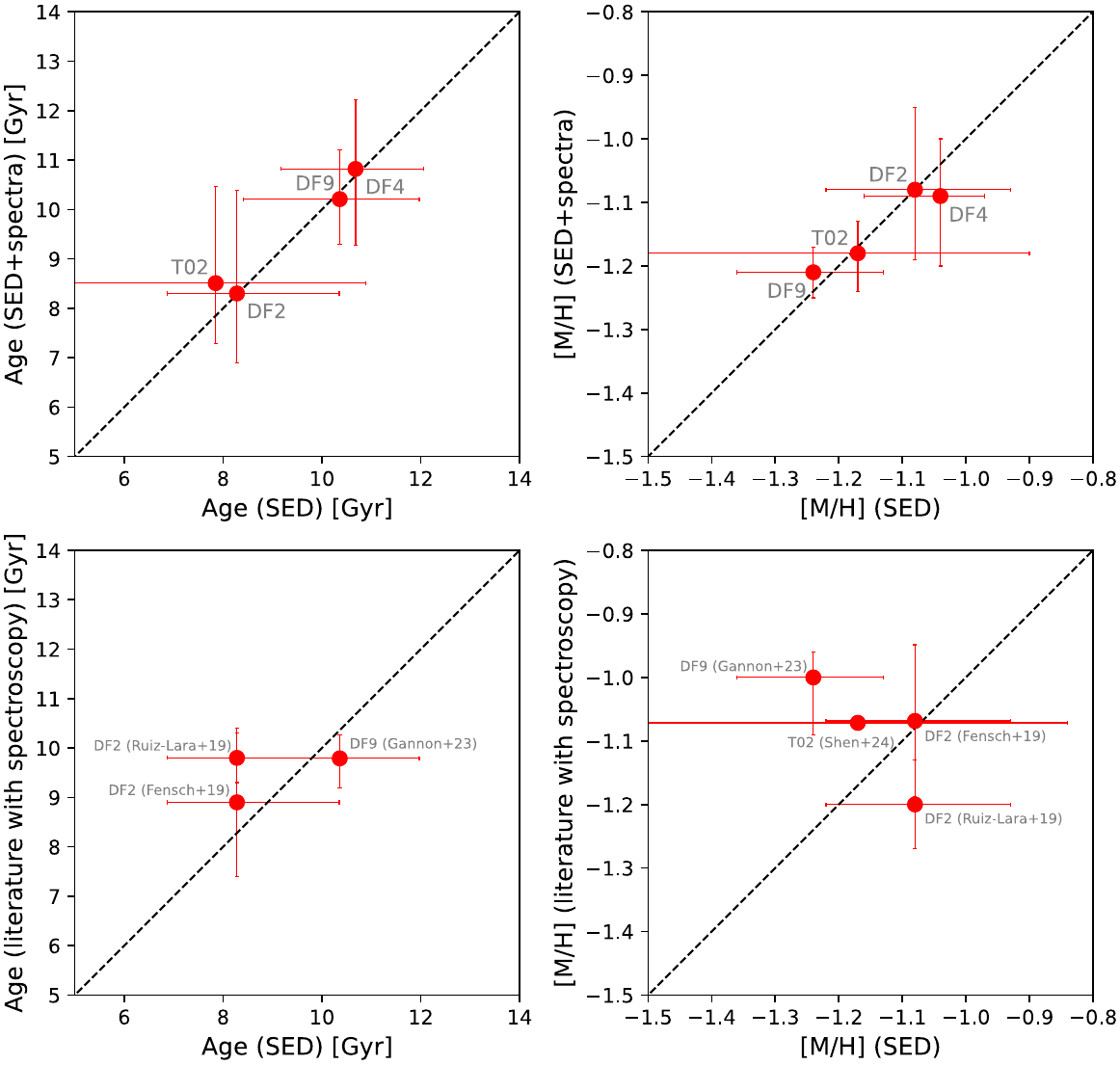}
\caption{{\it Top panels}: comparison of the mass-weighted age (left panel) and metallicity (right panel) between the {\tt PROSPECTOR} SED fitting and fitting with spectroscopy simultaneously for DF2, DF4, DF9, and T02. {\it Bottom panels}: comparison of the same two stellar population properties, but between our SED fitting and spectral spectroscopic results in the literature for DF2, DF9, and T02.}
\label{fig:stellar_pop_photospec}
\end{figure*}

Spectroscopic observations of quiescent, LSB dwarfs are challenging, with data available for four galaxies (three trail dwarfs and one non-trail dwarf) from the Keck Cosmic Web Imager (KCWI; \citealt{Morrissey2018}). These spectra allow us to carry out independent checks of the stellar populations results, while the photometry remains the main focus of the paper. Full details of the spectroscopic data and reduction are in the original papers discussed below, while the spectra of DF2, DF4, DF9, T02 and the best-fitting models from {\tt PROSPECTOR} are shown in Figure \ref{fig:df2_df4_df9_spec}.

The DF2 KCWI spectrum was presented in \citet{Danieli2019}, covering an aperture of $\sim$~10 arcsec ($\sim 0.5 R_{\rm e}$) radius. The medium slicer and BH3 grating were used, providing a spectral resolution of $\sim$~10,000 with a wavelength range of $\sim$~4800--5300~\AA. An offset pointing was used for sky subtraction. The S/N of the reduced spectrum is 20 \AA${}^{-1}$.

The DF4 spectrum was presented in \citet{Shen2023}, covering an aperture of $\sim$~7~arcsec ($\sim 0.5 R_{\rm e}$). The small slicer and BH2 grating were used, yielding a wavelength range of $\sim$~4000--4500~\AA\ and a spectral resolution of $\sim$~18,000. Similar to DF2, an offset sky pointing was used. The S/N of the reduced spectrum is 23 \AA${}^{-1}$.

The DF9 spectrum was presented in \citet{Gannon2023}, covering an aperture of $\sim$~8~arcsec ($\sim~0.8 R_{\rm e}$). The large slicer and BL grating were used, covering a wavelength range of 3554--5574~\AA\ with a spectral resolution of 785. The sky spectrum was ``on-chip,'' taken from the outer parts of the field of view, and the nuclear star cluster was also removed from the galaxy spectrum. The S/N of the reduced spectrum is 14 \AA${}^{-1}$.

The T02 spectrum was presented in \citet{Shen2024} (named as DFUWS-54), covering an aperture radius of $\sim$~10~arcsec ($\sim~0.7 R_{\rm e}$). The large slicer and BL grating were used, covering a wavelength range of $\sim$~3500--5500~\AA\ with a spectral resolution of $\sim$~900. Similar to DF9, the ``on-chip,'' sky spectrum was taken from the outer parts of the field of view, and the nuclear star cluster was also removed from the galaxy spectrum. The S/N of the reduced spectrum is 16 \AA${}^{-1}$.

First, we conduct spectrum-only fitting for these four galaxies. Our configurations in {\tt PROSPECTOR} are the same as mentioned in Section \ref{sec:prosp}. The DF2 and DF4 spectra have been smoothed to a spectral resolution of 2.5 \AA\ in order to match the resolution of the MILES templates used in the fitting. To match the low spectral resolution of DF9 and T02, we add velocity dispersion as an extra free parameter in the fitting. For DF4, DF9, and T02, the results are close to the broad-band SED fitting. \cite{Shen2023} excluded the blue end of the DF4 spectrum in their kinematics fitting, but we get similar results whether we include the blue end or not. 

In contrast, DF2 is found in the spectral fitting to be young and metal-rich, which is very different from the broad-band SED fitting results. We suspect that this discrepancy is related to the wavelength coverage of the spectrum of DF2, particularly because its left edge is close to the H$\beta$ absorption feature. When a continuum correction polynomial is applied (usually for any possible flux calibration issue), it is hard for the fitting software to know where the true continuum is on the blue side of H$\beta$, potentially leading to a poor reconstruction of the H$\beta$ line strength. 

To further explore the spectral range issue, we generated mock spectra with similar S/N to the four galaxies and extracted segments with different widths of spectral coverage. We find that the {\tt PROSPECTOR} results are sensitive to the correction polynomials, sometimes yielding large errors relative to the input properties, when the spectral ranges are small, as in the cases of DF2 and DF4 ($\sim$~500~\AA). We have also tried out {\tt pPXF} \citep{Cappellari2004} and found similar issues. Only if the wavelength coverage is wide enough to include many strong line features, such as the spectra of DF9 and T02, is there reliable recovery of the input parameters, as also shown by \cite{FerreMateu2023} in fitting KCWI spectra of other UDGs with similar wavelength ranges to our DF2 spectrum.

We also run {\tt PROSPECTOR} in a mode that fits the photometry and spectroscopy simultaneously. In this case, the photometry provides strong constraints on the continuum shape, and tempers the instabilities in the correction polynomials. The results are illustrated by Figure \ref{fig:stellar_pop_photospec} (top panels), where we find very similar results to the SED fitting\footnote{We note that the KCWI spectra cover only the inner regions instead of the entire galaxies. Even though none of these three galaxies exhibited significant color gradients in our measurements, there could be different gradients in age and metallicity that cancel each other out in color. Thus it is possible that an aperture difference could cause an apparent disagreement between spectral and SED fitting.}. With DF2 and DF4, the addition of spectroscopy does not really help with improving constraints on the stellar populations, owing to the limited spectral range. At least in these cases the SED-based results are confirmed as not inconsistent with the spectra (see fits in Figure~\ref{fig:df2_df4_df9_spec}). With DF9, the spectrum does reduce the uncertainties in age and particularly metallicity.
 
To further understand these comparisons, we generate mock data with similar S/N to those four galaxies, and run the fitting using photometry only, as well as using photometry and spectroscopy at the same time, while varying the spectral wavelength range. We find that at these S/N levels, broad-band SED fitting can robustly recover mass-weighted age and metallicity within the 1-$\sigma$ uncertainties. Adding spectra with a continuum S/N of 10--20 \AA${}^{-1}$ into the fitting slightly reduces the uncertainties in the results, but does not provide a big improvement. In comparison to pure broad-band SED fitting, this mixed method does not guarantee results closer to the true solutions, especially when the spectral range is not sufficiently broad.

Several galaxies in our sample have previous age and metallicity measurements obtained from spectroscopy, including DF2 \citep{vanDokkum2018b,Fensch2019a,Ruiz-Lara2019}, DF9 \citep{Gannon2023}, and T02 \citep{Shen2024}. The bottom panels in Figure \ref{fig:stellar_pop_photospec} show reasonable agreement between the literature and our results from SED fitting. The difference seen in the DF9 metallicity does not necessarily represent a discrepancy between spectroscopic and SED fitting. Our own fitting of the same spectrum also gives a different result from the literature, where {\tt pPXF} was used, but is very close to our SED fitting result -- reinforcing the challenge of systematics in such analyses and the importance of homogeneous studies with the same methods. Indeed, the top panel of Figure~\ref{fig:stellar_pop_photospec} illustrates the consistency obtained between spectral and SED fitting when using the same underlying models. As a final point of comparison, \cite{Ruiz-Lara2019} also recovered a non-parametric SFH for DF2. The results from our parametric SFH model are reasonably consistent in the two key parameters $t_M$ and $t_{90}$.

The study of \cite{Buzzo2022} was part of the inspiration for this work. They applied {\tt PROSPECTOR} SED fitting to a large sample of UDGs using optical to mid-infrared photometry. That work included DF2 and DF4 and returned very similar conclusions to ours in their relative ages, metallicities and stellar masses, albeit with larger uncertainties. We note, however, that \cite{Buzzo2022} reported $t_{\rm age}$ as the mass-weighted age, instead of the true $t_M$ calculated analytically within {\tt PROSPECTOR} using $t_{\rm age}$ and $\tau$ as input parameters. After correcting for this, their results indicate $t_M \sim$~4~Gyr for DF2 and $\sim$~7.5~Gyr for DF4, compared to our values of $t_M \sim$~8.5 and 10.5~Gyr, respectively. We have re-run their data-points using the methods in the current paper, including a different treatment of {\it WISE} upper limits and an updated version of {\tt PROSPECTOR}, and found $t_M \sim 8$~Gyr for both galaxies, again reasonably close to our results.

\begin{figure*}
\centering
\includegraphics[width=1\textwidth]{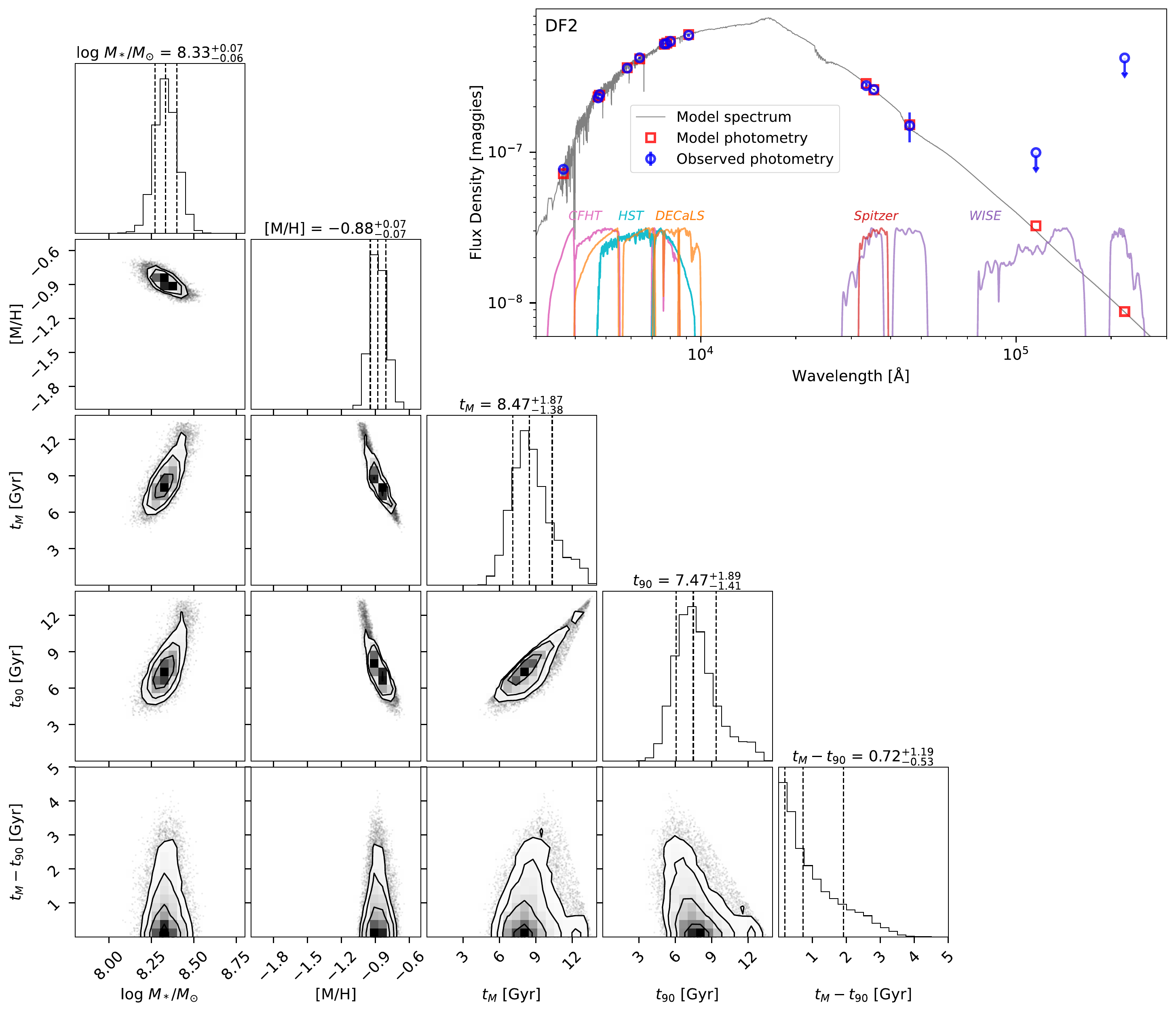}
\caption{As Figure \ref{fig:prospector_example}, but with the dust extinction $A_V=0$.}
\label{fig:prospector_df2_nodust}
\end{figure*}

Although we include dust as a free parameter, quiescent dwarf galaxies are conventionally thought to have almost no dust, and a finding of monochromatic GC populations in DF2 and DF4 provides evidence against dust in these galaxies \citep{vanDokkum2022b}. However, if dust is set to zero in our {\tt PROSPECTOR} fitting for the brightest galaxy in our sample (DF2), the metallicity returned is much higher than that reported in the literature from spectroscopy, shown in Figure~\ref{fig:prospector_df2_nodust}. Compared to Figure \ref{fig:prospector_example}, we mainly see a degeneracy between metallicity and dust inferences (which is not such a strong effect for the other bright galaxies in our sample). We suspect there could be a systematic problem with the photometry for DF2, perhaps related to having a more complicated stellar light distribution than a single S\'ersic model. Also, there might also be unidentified systematic problems with the stellar population synthesis models. For example, figure~6 from \citet{Johnson2021} shows similar amounts of dust recovered for Milky Way GCs that should be dust-free, which may be a by-product of model difficulties in perfectly reproducing colors of old, metal-poor populations (e.g.\ \citealt{Conroy2010}). As a compromise, we set a low dust prior ($0 < A_V < 0.2$ mag) to obtain more reliable inferences on age and metallicity.

We have examined all of the stellar population results from Section \ref{results_sed} using the no-dust models, and found the same conclusions. All of the internal and external checks above provide support for the reliability of our SED fitting results.

\end{document}